\documentclass[11pt,a4paper]{article}
\usepackage{jheppub}
\usepackage{array}
\usepackage{dsfont}
\usepackage{wrapfig}
\usepackage[small, it]{caption}
\usepackage{latexsym} 
\usepackage{amsmath, mathtools}
\usepackage{amssymb, amsthm}
\usepackage{graphicx}
\usepackage[english]{babel}
\usepackage{layout}
\usepackage{fancyhdr}

\usepackage{color}
\usepackage{psfrag}

\newcommand{\dif}{\mbox{d}} 
\newcommand{\ei}[1]{\,\mbox{#1}} 
\newcommand{\whizard}{\texttt{WHIZARD}}
\newcommand{\geneva}{\texttt{GenEvA}}
\newcommand{\pythia}{\texttt{PYTHIA}}

\title{An Analytic Initial-State Parton Shower}

\author[a]{Wolfgang~Kilian,}
\author[b]{J\"urgen~Reuter,}
\author[b]{Sebastian~Schmidt}
\author[b]{and Daniel~Wiesler}

\affiliation[a]{Department Physik, Universit\"at Siegen, D--57068
  Siegen, Germany} 
\affiliation[b]{DESY Theory Group, Notkestrasse 85, D--22607 Hamburg,
  Germany}

\emailAdd{kilian@physik.uni-siegen.de}
\emailAdd{juergen.reuter@desy.de}
\emailAdd{sebastian.t.schmidt@desy.de}
\emailAdd{daniel.wiesler@desy.de}

\abstract{We present a new algorithm for an analytic parton shower. While the
algorithm for the final-state shower has been known in the
literature, the construction of an initial-state shower along these
lines is new. The aim is to have a parton shower algorithm
for which the full analytic form of the probability distribution for
all branchings is known. For these parton shower algorithms it is
therefore possible to calculate the probability for a given event to
be generated, providing the potential to reweight the event after the
simulation. 
We develop the algorithm for this shower including scale choices and
angular ordering. Merging to matrix elements is used to describe
high-energy tails of distributions correctly. Finally, we compare our results with those of other
parton showers and with experimental data from LEP, Tevatron and
LHC.
}


\arxivnumber{1112.1039}

\begin{document}
\maketitle
   
\newpage


\section{Introduction}
Parton Showers are an indispensable part in the simulation of hadronic collisions in today's high-energy colliders, like the now-running LHC at CERN. A precise simulation of these collisions demands a coordinated interplay of such diverse elements as the calculation of hard matrix elements, the simulation of parton showers, a matching procedure to combine matrix elements and parton shower, the modeling of the underlying event and the hadronization and the simulation of the detector response. The main task of parton showers in this framework is to describe collinear and soft emissions off incoming or outgoing partons from the hard matrix element, thereby affecting the jet substructure and possibly increasing the number of resolved jets. The parton shower implementation most commonly used is the one in \pythia\ \cite{PYTHIA64,Sjostrand:2007gs}. Alternatively, programs such as Herwig++ \cite{Bahr:2008pv} and \texttt{SHERPA} \cite{Gleisberg:2008ta} implement parton shower algorithms as components for their event generation frameworks. Moreover programs that are tailored exclusively to parton showers and work as plug-ins to event generators are available, the most prominent being Vincia \cite{Giele:2007di}.

Recently it was shown that the formulation of parton showers can be reproduced by the soft-collinear effective field theory \cite{bauer-2007-76}. This led to the formulation of so-called \emph{analytic parton showers} \cite{Gaining}.
In this work, we intend to use the analytic parton shower for event generation. We extend the analytic final-state parton shower presented by Bauer et al. \cite{Geneva2} and develop a new analytic parton shower for initial-state radiation. During this work, we implemented both these parton showers in the event generator \whizard\ \cite{Kilian:2007gr}.
We succeeded in reproducing event shape distributions simulated using \pythia's parton shower, as well as distributions measured at LEP, Tevatron and LHC very well, given that very little tuning was done.

Section \ref{sec:theoryPS} describes the theory of the parton shower derived from the analytic approach, as well as the extensions and improvements implemented in the final-state parton shower and the newly implemented parton shower for the initial state. Section \ref{sec:extension} describes the steps to be taken for a comparison of the implementation of our algorithms to data and other showers. These include the implementation of a MLM-type matching, described section \ref{sec:matching}, an interface to an external hadronization routine, in our case \pythia, in section \ref{sec:hadronization} and the handling of beam-remnants, described in section \ref{sec:beamremnants}.
In section \ref{sec:results}, we show distributions obtained using our shower or \pythia's shower for various event and jet shapes. Furthermore we compare those to the respective measurements at LEP, Tevatron and LHC.
Finally, in section \ref{sec:conclusion} we summarize our findings, conclude and give an outlook on future developments.

\section{The Analytic Parton Shower - Introduction and Algorithm}
\label{sec:theoryPS}

\subsection{Concept of Parton Showers}

Parton showers are commonly formulated using branchings of one
particle into two, which can either be considered as one parton
splitting into two partons or one parton emitting a new
parton\footnote{The exception to this are the showers based on the
  dipole/antenna picture and their formulation using two to three
  splittings, originally developed in \cite{Gustafson:1987rq}.}. The
central entity of the parton shower is the Sudakov form factor
$\Delta$ --  originally described in \cite{MR0077427} --  its simplest form is given by
\begin{equation}
	\Delta( t_1, t_2) = \exp \left[ \int\limits_{t_1}^{t_2} \dif t \int\limits_{z_-}^{z_+} \dif z \frac{\alpha_s}{2 \pi t} P(z) \right] 
\end{equation}
giving the probability for a parton to evolve from scale $t_2$ to $t_1$ without emitting a further parton. Therein, the variable $z$ describes the relation of the two partons after the branching, the most common choice is to take $z$ to be the ratio of one parton's energy after the branching and the parton's energy prior to the branching. The functions $P(z)$ are called splitting functions. They describe the probabilities for the respective branchings and can be inferred from approximate calculations of matrix elements. The variable $t$ is called the scale. There is a certain ambiguity how to define the scale in an implementation. Here we use the virtuality that is defined as the square of the four-momentum $t=p^2$.
The simulation of the parton shower is an evolution in the scale. For
final-state radiation, the branchings that occur after the hard
interaction, the evolution is from a scale corresponding to the hard
interaction, $t \sim \hat{s}$, down to a cut-off scale $t=t_{cut}$,
that symbolizes the transition to the non-perturbative physics
encapsulated in the hadronization. For initial-state
radiation\footnote{For now, only initial-state radiation for partons
  stemming from hadrons are considered.}, branchings that appear
before the hard interaction, the evolution is from a cut-off
$t=-t_{cut}$, representing the factorization scale and thus the parton
density functions, down to a scale corresponding to the negative of
the center of mass energy, $t \sim - \hat{s}$\footnote{This treatment
  is only correct for $s$-channel resonance decays, like the processes
  considered in this paper. For general processes a more sophisticated
  ansatz will be implemented in the future.}. In the implementation, this evolution, that corresponds to an evolution in physical time, is replaced by an evolution starting at the hard interaction and ending at the cut-off scale. This is the so-called backwards evolution, originally described first in \cite{Sjostrand:1985xi}, its most prominent consequence being the appearance of parton density functions in the Sudakov factor.

\subsection{The Analytic Parton Shower}

The parton shower is a well-defined approximation to the full matrix element. Therefore it should be preferable to be able to reconstruct the matrix element from the parton shower. 
In common parton shower algorithms, this ability is lost due to the formulation of the parton shower as a Markov chain in such a way, that branchings that fail to respect correct kinematics can be produced and are subsequently rejected or manually modified to respect momentum conservation. It is these branchings that prevent the probability for a branching to be calculated analytically after the branching is generated. 
Therefore in developing an analytic parton shower, care was taken to avoid branchings that need to be rejected or manually modified, thereby preserving the ability to reconstruct the matrix element. The two main changes are the simultaneous simulation of the branchings of sisters and replacing the splitting variable $z$, that normally is the ratio of the first daughter's energy or light-cone momentum to the mothers' energy or light-cone momentum.

\begin{figure}[ht]
\centering
\psfrag{t0}[l][][1][0] {$t_a$}
\psfrag{t2}[l][][1][0] {$t_b$}
\psfrag{t3}[l][][1][0] {$t_c$}
\includegraphics[height=45mm]{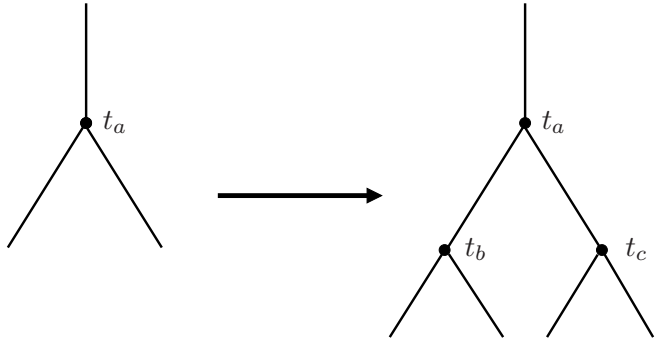}
\caption{Schematic view of a double branching. Before the double branching (left): A parton has branched at a scale $t_a$ into two on-shell daughter partons. After the simulation of the double branching the branching scales $t_b, t_c$ for the daughter partons are known. In case the daughters branch themselves the needed values are generated as well. The case in which both daughters branch, $t_b>t_{cut}, t_c > t_{cut}$, is shown on the right. For the next step the double branchings of the two daughter partons will be simulated, the branchings at $t_b$ and $t_c$, respectively, replace the parton branching at $t_a$ on the left.}
\label{fig:doublebranchingfsr}
\end{figure}

The first modification is to replace the simulation of individual
branchings by the simulation of so-called double branchings. A double
branching consists of the simultaneous branching (or no-branching) of
the two daughter-partons of one parton
(cf. figure~\ref{fig:doublebranchingfsr}). So instead of taking one 
parton $a$ and letting it branch into two partons, $a \rightarrow bc$,
an existing branching $a \rightarrow bc$ is replaced by the double
branching $a\rightarrow bc \rightarrow defg$ with the new partons
$d,e,f,g$ in case both partons $b$ and $c$ branch. The corresponding
situations where one or both of the daughters do not branch are also
taken into account. The advantage is that the
energy-conservation\footnote{The following equation can best be
  understood in the rest frame of the mother. Then $\sqrt{t_a}$ is the
  mother's mass and energy and trivially the sum of the daughters'
  masses $\sqrt{t_b} + \sqrt{t_c}$ has to be less.} 
\begin{equation}
	\sqrt{t_a} \geq \sqrt{t_b} + \sqrt{t_c}
\end{equation}
can be included in the generation of the branchings, avoiding the production of complicated interconnections between different single branchings. The sequence of steps is then
\begin{itemize}
\item Pick a branch with unprocessed daughters $b$ and $c$.
\item Generate $\{t_b, v_b\}$ and $\{t_c, v_c\}$ for both daughters independently with the probability given by the single branching probability. ($v_i$ stands for the values needed to describe the branching apart from the virtuality $t_i$, like the opening angle $\cos\vartheta$, the azimuthal angle $\phi$ and the type of the daughter parton.) 
\item Keep the branch of the daughter with the higher scale $t_{max} = \max\left(t_b, t_c\right)$. Discard the branching of the other daughter.
\item Determine new values for the other daughter with the maximum scale set to $t_* = \min\left[ t_{max}, \left(\sqrt{t_a} - \sqrt{t_{max}} \right)^2 \right]$.
\end{itemize}
For the different cases, the double branch probabilities can be constructed from the single branching probabilities \cite{Gaining} $\mathcal{P}^{br}_i(t_i,v_i)$ and Sudakov factors $\Delta_i(t_a,t)$ for a branching at the scale $t_i$ and the remaining values $v_i$ and the probability $\mathcal{P}^{nb}_i$ for no branching above the cut-off. The double branch probabilities for the case in which both daughters branch is
\begin{eqnarray}
	\mathcal{P}^{br,br} (t_b, v_b, t_c, v_c) &=& \theta(t_b - t_c) \mathcal{P}_b^{br} (t_b, v_b) \Delta_c(t_a, t_b) \\
	&& \mathcal{P}_c^{br} (t_c, v_c; t_a=t_*) \nonumber \\
	&+& \theta(t_c - t_b) \mathcal{P}_c^{br} (t_c, v_c) \Delta_b(t_a, t_c) \nonumber \\ 
	&& \mathcal{P}_b^{br} (t_b, v_b; t_a=t_*) \nonumber,
\end{eqnarray}
while in the case that only one daughter branches
\begin{eqnarray}
	\mathcal{P}^{br,nb} (t_b, v_b) = \mathcal{P}_b^{br} (t_b, v_b) \Delta_c(t_a, t_b) \Delta_c(t_*, t_{cut}) \\
	\mathcal{P}^{nb,br} (t_c, v_c) = \mathcal{P}_c^{br} (t_c, v_c) \Delta_b(t_a, t_c) \Delta_b(t_*, t_{cut}),
\end{eqnarray}
and in the case no parton branches
\begin{equation}
	\mathcal{P}^{nb,nb} = \Delta_b(t_a, t_{cut}) \Delta_c(t_a, t_{cut}).
\end{equation}
Taking all different combinations into account, the double branch probability can be composed in the following way
\begin{eqnarray}
	\mathcal{P} (t_b, v_b, t_c, v_c) &=& \mathcal{P}^{br,br} (t_b, v_b, t_c, v_c) \nonumber \\
	&+& \mathcal{P}^{br,nb} (t_b, v_b) \quad \delta(t_c) \nonumber \\
	&+& \mathcal{P}^{nb,br} (t_c, v_c) \quad \delta(t_b) \nonumber \\
	&+& \mathcal{P}^{nb,nb} \quad \quad \delta(t_b) \delta(t_c). \label{eq:doublebranchprobabilityFSR}
\end{eqnarray}

The second step is replacing the kinematic ratio $z$ with the angle $\cos\theta$ in the mother's rest frame between the momentum of the first daughter and the boost axis. This leads to simple phase space limits
\begin{displaymath}
	-1 \leq \cos \theta \leq 1.
\end{displaymath}
There is a direct correspondence between the $\cos\theta$ angle and the energy splitting $z$ \cite{Gaining} as a function of the masses of the daughters $t_b$ and $t_c$:
\begin{equation}
	z = \frac12 \left[ 1 + \frac{t_b}{t_a} - \frac{t_c}{t_a} + \beta_a \cos\theta_a \lambda(t_a, t_b, t_c) \right]
\end{equation}
with the boost $\beta_a$ and the phase space factor $\lambda$:
\begin{displaymath}
	\beta_a = \sqrt{1 - \frac{t_a}{E_a^2}} \qquad \mbox{and} \qquad \lambda(t_a, t_b, t_c) = \frac1{t_a} \sqrt{ (t_a - t_b - t_c)^2 - 4 t_b t_c}
\end{displaymath}
The important distinction between common and analytic parton showers is that in the analytic parton shower every branching is generated with a calculable probability. Every source for vetoing branchings where the probability for the veto cannot be calculated has therefore been avoided.

\subsection{Improved Analytic Final State Parton Shower}

The \geneva\ framework \cite{Geneva1,Geneva2} is an event generation framework designed to combine matrix elements and parton showers during event generation. It uses its parton shower to distribute events over phase space, in order to reweight them to a corrected distribution later.
Therefore only a simplified implementation of parton showers was included in the framework, as the reweighting would later reintroduce the correct distributions. 
We, on the other hand, will use the analytic parton shower to generate physical events and cannot defer anything to a reweighting procedure, we are therefore forced to implement the full theory of parton showers. The two main simplifications made in the \geneva\ framework are the omission of the running of the coupling constant and the omittance of color coherence. Our extensions to the parton shower are as follows.

The running of the coupling constant $\alpha_S(Q^2)$ was implemented, the inclusion is straight-forward.
The coupling was chosen to be
\begin{equation}
	\alpha_S = \alpha_S \left( z(1-z)Q^2 \right) = \alpha_S(z(\theta), Q^2)
\end{equation}
in agreement with most parton shower generators.

As color coherence is approximated by demanding that the angles of subsequent emissions decrease -- this is known as angular ordering -- the resulting phase space cuts have to be implemented in the parton shower. The opening angle\footnote{not to be confused with the angle $\cos\theta$ used in the description of branchings} $\cos\vartheta$ is given by
\begin{equation}
	\cos \vartheta = 1 - \frac{t}{2z(1-z)E^2}
\end{equation}
in the approximation for massless children. Using $z(1-z) \leq \frac14$ this can be used to give a cut on the scale of a next branching
\begin{equation}
	t \leq E^2 \frac{1 - \cos\vartheta_{cut}}{2}
\end{equation}
for the branching to have an opening angle less than $\cos\vartheta_{cut}$. An additional cut on $z$ \cite{Gaining} is necessary
\begin{equation}
	\left|z - \frac12 \right| \leq \frac{\beta}{2} \sqrt{1 - \frac{t}{\beta^2 E^2} \frac{1+\cos\vartheta_{cut}}{1-\cos\vartheta_{cut}}}.
\end{equation}
With these phase space cuts angular ordering is enforced in the approximation of massless daughter partons. However the inclusion of this constraint demands keeping track of the used energy $E$ and the used angle $\cos\vartheta_{cut}$ either by explicitly storing their values for every branching or by using a distinct rule to calculate them for every branching.

As a minor extension we allow for parton masses, although these are only taken into account when distributing momenta, the splitting functions are still taken for massless daughter partons.

\subsection{Introducing the Analytic Initial State Parton Shower}

\begin{figure}
\begin{center}
\psfrag{t2}[l][][1][0] {$t_b$}
\psfrag{t1}[l][][1][0] {$t_a, z_a$}
\psfrag{t3}[l][][1][0] {$t_c (v_c)$}
\includegraphics[height=50mm]{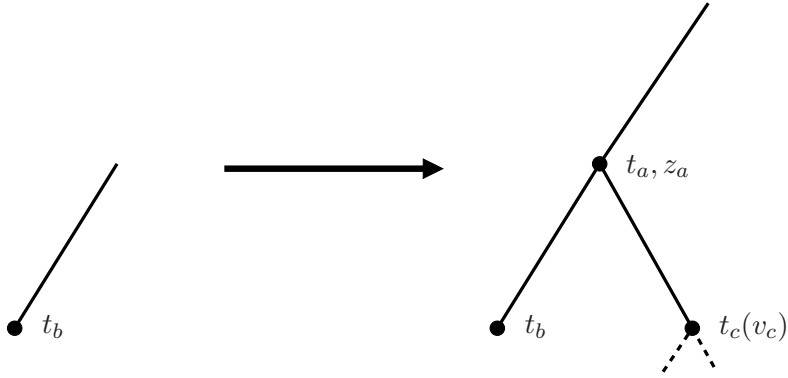}
\end{center}
\caption{Schematic view of a double branching in ISR: Before the double branching (left): At the scale $t_b$ a parton $b$ exists. For this parton, the scale $t_a$ of the branching that produced this parton, the corresponding energy ratio $z_a$ and the scale $t_c$ where the emitted parton $c$ branches, and if necessary the remaining quantities ($v_c$) are simulated (on the right).}
\label{fig:doublebranchingisr}
\end{figure}

For physics at the LHC, a parton shower has to be able to describe both, initial and final-state radiation. We therefore implement an initial-state parton shower satisfying the requirement of analyticity analogously to the parton shower for the final state.

The changes applied to the final-state shower cannot be transferred to the initial-state shower\footnote{Due to the negative virtualities all momenta and energies would be imaginary in the mothers restframe.}. A different set of changes is needed to reformulate the initial-state parton shower in order to fulfill the demand of analyticity. The known Sudakov factor for initial-state radiation is
\begin{align}
	\Delta_b^{ISR}(t_a, t_b) = \exp \left[ - \int\limits_{|t_a|}^{|t_{b}|} \dif t' \int\limits_0^1 \frac{\dif z}{z} \frac{\alpha_S}{2 \pi t'} \left. \left. \sum\limits_{a,c} \right( P_{a\rightarrow bc}(z) + P_{a\rightarrow cb}(z) \right) \frac{f_a\left(\frac{x_b}{z}, t' \right)}{f_b(x_b,t')} \right]
\end{align}
with the splitting function $P_{a\rightarrow bc}(z)$ for a parton of type $a$ branching into two partons of type $b$ and $c$ and the parton density functions $f_a(x,t)$.
The conservation of momentum can be enforced by explicitly vetoing momentum conservation-violating branchings directly in the Sudakov factor. To do this, the branching of the mother parton and the branching of the emitted parton have to be simulated simultaneously, cf. figure \ref{fig:doublebranchingisr}. The simulated branching therefore effectively becomes a $1\rightarrow 2$ (if the emitted parton does not branch) or a $1\rightarrow 3$ (if the emitted parton branches) branching. The Sudakov factor that takes the emitted parton's branching into account can be written in the form
\begin{eqnarray}
	&& \Delta_b^{ISR}(t_a, t_b) = \exp \Biggl[ - \int\limits_{|t_a|}^{|t_{b}|} \dif t' \int\limits_0^1 \frac{\dif z}{z} \frac{\alpha_S}{2 \pi t'} \sum\limits_{a,c} \int\limits_{0}^{t'} \dif t_c \, \mathcal{P}_c(t_c | -t', z)  \nonumber \\ 
	&& \qquad \quad \bar{\Theta}\left(-t', t_b, t_c , z_a, E_a\right) \biggr( P_{a\rightarrow bc}(z) + P_{a\rightarrow cb}(z) \biggr) \frac{f_a\left(\frac{x_b}{z}, t' \right)}{f_b(x_b,t')} \Biggr]
	\label{eq:sudalgo}
\end{eqnarray}
with the veto function
\begin{equation}
	\bar{\Theta}\bigl(t_a, t_b, t_c , z_a, E_a\bigr) = \Theta\bigl( |\vec{p}_b| + |\vec{p}_c| -|\vec{p}_a| \bigr) \cdot \Theta \bigl(|\vec{p}_a| - \left| |\vec{p}_b| -|\vec{p}_c| \right| \bigr),
\end{equation}
and the one parton branching distribution function for the emitted parton $c$
\begin{equation}
	\mathcal{P}_c(t_c | -t', z),
\end{equation}
giving the probability distribution for the branching of the emitted timelike parton as a function of the branching this parton was produced, described by $-t'$ and $z$. 
The veto function ensures that the three partons $a,b,c$ can be combined in a branching that conserves momentum by enforcing the triangle inequality. By adding more terms it can also be used to impose cuts for angular ordering or a minimum energy for the emitted timelike parton. If the emitted parton branches, its final-state parton shower can now be simulated further by the use of the known double branching probabilities from the analytic final-state radiation.

However there is a slight difference in the interpretation of the known one-branching Sudakov factor as used for example in \pythia\ and the supplemented one in equation \eqref{eq:sudalgo}. In the former, the probability for a branching is independent of the available allowed branchings of the emitted parton, while in the latter the probability for a branching is reduced when the emitted parton has a restricted phase space for branchings. Therefore the supplemented Sudakov factor rather resembles a conditional probability.

Using these prescriptions, double branch probability distributions can be formulated, analogously to the ones formulated for final-state radiation. The probability for no earlier branching, the parton being directly emitted by the hadron and therefore being on-shell, $t_a \rightarrow m_a^2$, consists of the Sudakov factor $\Delta_b^{ISR} ( -t_{cut}, t_b)$ and a $\delta$-distribution forcing the parton to be on-shell and thus can be formulated in the form
\begin{equation}
	\mathcal{P}^{nb}_b ( t_a; t_b, t_{cut} ) = \Delta_b^{ISR} ( -t_{cut}, t_b) \delta\left(t_a - m_a^2 \right).
\end{equation}
In case an earlier branching is found, the common single branch probability would be
\begin{eqnarray}
	 \mathcal{P}_{a \rightarrow bc} ( t_a, z_a; t_b, t_{cut} ) &=& \frac{\alpha_S}{2 \pi t_a} \frac{1}{z_a} P_{a\rightarrow bc}(z_a) \frac{f_a\left(x_a, t_a \right)}{f_b\left(x_b,t_a\right)} \nonumber \\ 
	&& \cdot \; \Delta_b^{ISR} (t_a, t_b) \Theta\left( t_a - t_b \right) \Theta\left(-t_a - t_{cut}\right)
\end{eqnarray}
with the Sudakov factor $\Delta_b^{ISR} (t_a, t_b)$, a relative weight, given by the ratio of parton density functions, $\frac{f_a\left(x_a, t_a \right)}{f_b\left(x_b,t_a\right)}$, the probability for the branching itself, $\frac{\alpha_S}{2 \pi t_a} \frac{1}{z_a} P_{a\rightarrow bc}(z_a)$ and two step functions that force the parton to be in the correct range of virtuality.

For the transition to analytic showers, a dependence on the scale of the emitted parton $t_c$ is introduced. Thus two different cases have to be considered. In the case of the emitted parton not branching further, the corresponding probability distribution is supplemented by the no-branching-probability $\mathcal{P}_c^{nb}$ for the emitted parton $c$ and the veto function $\bar{\Theta}$. It can be written in the way
\begin{eqnarray}
	 \lefteqn{\mathcal{P}^{br,nb}_{a \rightarrow bc} ( t_a, t_c, z_a; t_b, t_{cut} )} \nonumber \\
	&=& \frac{\alpha_S}{2 \pi t_a} \bar{\Theta}\left(t_a, t_b, t_c , z_a, E_a\right) \frac{1}{z_a} P_{a\rightarrow bc}(z_a) \frac{f_a\left(x_a, t_a \right)}{f_b\left(x_b,t_a\right)} \nonumber \\ 
	&& \cdot \; \Delta_b^{ISR} (t_a, t_b) \Theta\left( t_a - t_b \right) \Theta\left(-t_a - t_{cut}\right) \mathcal{P}_c^{nb} \left(t_c; |t_a|, t_{cut}\right).
	\label{eq:WbrnbISR}
\end{eqnarray}
In the case, the emitted parton undergoes another branching, the distribution is supplemented by the single branch probability for the emitted parton $\mathcal{P}_{c\rightarrow de}^{br}$ and has the form
\begin{eqnarray}
	\lefteqn{\mathcal{P}^{br,br}_{a \rightarrow bc \rightarrow bde} ( t_a, t_c, z_a, v_c; t_b, t_{cut} )} \nonumber \\
	&=& \frac{\alpha_S}{2 \pi t_a} \bar{\Theta}\left(t_a, t_b, t_c , z_a, E_a\right)  \frac{1}{z_a} P_{a\rightarrow bc}(z_a) \frac{f_a\left(x_a, t_a \right)}{f_b\left(x_b,t_a\right)} \nonumber \\
	&& \cdot \; \Delta_b^{ISR} (t_a, t_b) \Theta\left( t_a - t_b \right)\Theta\left(-t_a - t_{cut}\right) \mathcal{P}_{c\rightarrow de}^{br} \left(t_c, v_c; |t_a|, t_{cut}\right).
	\label{eq:WbrbrISR}
\end{eqnarray}

Using these expressions the probability distribution for the scale $t_a$ can therefore be written analogously to equation \eqref{eq:doublebranchprobabilityFSR} in the form
\begin{eqnarray}
	\lefteqn{\mathcal{P}_b(t_a; t_b, t_{cut}) = \mathcal{P}^{nb}_b ( t_a; t_b, t_{cut} )} \nonumber \\
	&+& \sum\limits_{a,c} \int\dif z_a \int\dif t_c \, \mathcal{P}^{br,nb}_{a \rightarrow bc} ( t_a, t_c, z_a; t_b, t_{cut} ) \nonumber \\
	&+& \sum\limits_{a,c} \int\dif z_a \int\dif t_c \, \mathcal{P}^{br,nb}_{a \rightarrow cb} ( t_a, t_c, z_a; t_b, t_{cut} ) \nonumber \\
	&+& \sum\limits_{a,c,d,e} \int\dif z_a \int\dif t_c \int\dif v_c \, \mathcal{P}^{br,br}_{a \rightarrow bc \rightarrow bde} ( t_a, t_c, z_a, v_c; t_b, t_{cut} ) \nonumber \\
	&+& \sum\limits_{a,c,d,e} \int\dif z_a \int\dif t_c \int\dif v_c \, \mathcal{P}^{br,br}_{a \rightarrow cb \rightarrow deb} ( t_a, t_c, z_a, v_c; t_b, t_{cut} ).
	\label{eq:Wb}
\end{eqnarray}
The probability distributions for the parton species and the energy fractions $z$ follow directly from this equation.

One aspect of initial-state parton showers that is a rather critical technical point, is the assignment of momenta for the first branchings in the initial state. By the first branchings we mean the respective branchings closest to the hard interaction for the two incoming partons in the matrix element. As these partons are on the mass-shell and often assumed to be massless, any branching would be kinematically forbidden. Therefore the partons have to be set off-shell in order to allow for kinematically allowed branchings. This is done by simultaneously scaling the partons' momenta, until the four-momentum squared reaches the scale of the first branching,
\begin{displaymath}
	t = p^2 = E^2 - \vec{p}^2 = t_{first} \ll 0.
\end{displaymath}
The distribution of $t_{first}$ is obtained by solving a Sudakov factor similar to the one given in equation (\ref{eq:sudalgo}), but with the terms corresponding to the emitted parton removed,
\begin{eqnarray}
	&& \Delta_b^{ISR}(t_{first}, t_b) = \exp \Biggl[ - \int\limits_{|t_a|}^{|t_{b}|} \dif t' \int\limits_0^1 \frac{\dif z}{z} \frac{\alpha_S}{2 \pi t'} \sum\limits_{a,c} \qquad  \nonumber \\ 
	&& \qquad \qquad \qquad \biggr( P_{a\rightarrow bc}(z) + P_{a\rightarrow cb}(z) \biggr) \frac{f_a\left(\frac{x_b}{z}, t' \right)}{f_b(x_b,t')} \Biggr].
	\label{eq:sudalgofirst}
\end{eqnarray}
Thus the initial-state parton shower is not started from the two partons in the initial state of the matrix element, but from copies of them that have their momenta assigned in the following way:
\begin{eqnarray}
  t_1 = t_{first\,1} &\qquad&  t_2 = t_{first\,2} \\
  E_1 = \frac{ \hat{s} + t_1 - t_2 }{4 \sqrt{\hat{s}}} &\qquad& E_2 = \frac{\sqrt{\hat{s}}}2 - E_1 \\
  |\vec{p}_1| = |\vec{p}_2| = \sqrt{ E_1^2 - t_1 } &\qquad& \vec{p}_1 = -\vec{p}_2
\end{eqnarray}
By doing so, both partons are set off-shell so that branchings are kinematically allowed, while conserving the total energy and momentum. Another possibility would be to enlarge the three-momenta so that the scales are equal to the negative partonic center-of-mass energy, $t \rightarrow -\hat{s}^2$, and then start the shower from there, but this starting configuration has the disadvantage that the three-momenta of the initial partons tend to be very large, so that it becomes very hard to find kinematically allowed branchings.

\section{Prerequisites for a realistic description}
\label{sec:extension}

In this section we discuss the technical prerequisites for a realistic
implementation of our analytic parton shower algorithm and the
preparations needed to compare results with results from other parton showers 
and with experimental data. To do so, we chose to implement the shower
algorithm within the event generator
\whizard~\cite{Kilian:2007gr} which contains
highly optimized (tree-level) matrix elements by the matrix-element
generator \texttt{O'Mega}~\cite{Moretti:2001zz}, a very efficient
phase-space parametrization and a multi-channel adaptive Monte-Carlo
integration~\cite{Ohl:1998jn}. \whizard\ has been developed and
has found a wide range of application to lepton collider physics 
(cf.~e.g.~\cite{Ohl:2004tn,Beyer:2006hx,Kalinowski:2008fk}). The program has
been completely recast for hadron collider physics and been 
successfully applied to BSM and jet physics
(cf.~e.g.~\cite{Hagiwara:2005wg,Kilian:2006eh,Alboteanu:2008my,Christensen:2010wz,Reuter:2010nx}). 
Several steps to include NLO corrections in a semi-automatic way have
been 
undertaken~\cite{Kilian:2006cj,Robens:2008sa,Binoth:2009rv,Greiner:2011mp}.

To make contact with experimental
distributions, one has to cover the whole of phase space to access the
high-energy tails of distributions. One possibility is using the so-called
\emph{power-shower} concept where one artificially opens up more phase
space than physically available to generate hard and/or non-collinear
jets from the parton shower. We decided not to use this concept, but to implement a matching
procedure of matrix elements with explicit additional jets with the
showered Born process. This is done in the first part of this
section. In the second part we discuss our treatment of hadronization to compare a realistic event simulation with experimental data, while in the last
section we explain how beam remnants are dealt with in our framework.

The extensions presented in this section will become publicly available in all future releases from \whizard\ 2.1 on.

\subsection{Matching}
\label{sec:matching}

A matching procedure is a procedure to combine the description of up to a certain number of multiple, widely separated jets by the matrix element and the description of possible additional jets and the substructure of the jets by a parton shower. The main approaches for matching to leading order calculations are the CKKW \cite{Catani:2001cc}, CKKW-L \cite{Lonnblad:2001iq} and MLM \cite{MLM,Mangano:2001xp} schemes, for a general overview see \cite{Alwall:2007fs,Hoche:2006ph,Lavesson:2007uu}. In the process of implementing the analytic parton shower, we also implemented a matching procedure according to the principles of the MLM approach in \whizard\ with the use of the \texttt{KTCLUS} clustering package \cite{Catani1993187}.

The steps as implemented in \whizard\ are:
\begin{enumerate}
\item The cross-sections for the main process and processes with up to $N$ additional partons in the hard matrix-element are calculated. The phase space has to satisfy the additional cuts enforced by the matching procedure
\begin{equation}
	p_T > p_{T\,min} , \qquad |\eta| < |\eta_{max}|, \qquad \Delta R_{jj} > R_{min}
\end{equation}
with the transverse momentum $p_T$, the pseudo-rapidity $\eta$ and the $\eta-\phi$-distance between two jets $\Delta R_{jj}$. The values $p_{T\,min}$, $\eta_{max}$ and $R_{min}$ can be set in the \whizard\ input file.
\item According to the relative probability $P(i)$ given by the relative size of the corresponding cross-sections,
\begin{equation}
	P(i) = \frac{\sigma_i}{\sum_{j} \sigma_j}
\end{equation}
a matrix-element event with $i$ additional partons is generated.
\item These events are then showered with the analytic shower\footnote{As an alternative, one could also use the \pythia\ shower.}.
\item After the shower evolution, a $k_T$-clustering jet algorithm \cite{Catani1993187} is applied to the showered, but not yet hadronized event, taking only colored partons with a pseudo-rapidity $|\eta|<\eta_{max\,clus}$ into account. Jets are defined by a minimum jet-jet separation $y_{cut}$\footnote{see section \ref{sec:jetrates} for a short introduction to jet clustering}
\begin{eqnarray}
	\eta_{max\,clus} &=& \eta_{clusfactor} * \eta_{max} \label{eq:etaclus}\\
	y_{cut} &=& \Big[p_{T\, min} + \mbox{max}\left(E_{TclusminE}, \right.   \nonumber \\
	& & \left. \qquad E_{T\,clusfactor}*p_{T\,min} \right) \Big]^2 \label{eq:ETclus}
\end{eqnarray}
The factors and hence the clustering variables can be varied as part of the systematics assessment, the defaults for these factors are chosen to be 1.
\item If the jet algorithm in the matching procedure undershoots the number
of matrix element jets the event is discarded. When the event after the jet merging overshoots the number
of matrix element jets, the event is rejected as well, unless the
number of matrix element jets is equal to the maximum number of matrix element jets. In
that case the scale $y_{cut}$ is adapted such that the number of
reconstructed jets is reduced to the number of matrix element jets,
i.e. the jet resolution is lowered accordingly.
\item Then it is tested if the reconstructed jets match the matrix element partons. This is done in an iterative way: The clustering is reapplied to a set consisting of the reconstructed jets and one matrix element parton. If this additional parton leads to an additional jet above the scale $y_{cut}$, the parton is assumed not to be matched to any of the reconstructed jets and the event is discarded. Otherwise the matched reconstructed jet is removed from the set and this step is repeated for the next matrix element parton. If and only ifall matrix element partons can be matched in this way, the event is accepted. 
\item The remaining steps of event generation, like multiple interactions, hadronization, and pile-up, can then be applied. 
\end{enumerate}

\subsection{Hadronization}
\label{sec:hadronization}

Hadronization of the generated events was delegated to \pythia. An interface between \whizard\ and \pythia\ was written for this purpose.
A detailed description of the interface will be given in the
\whizard\ manual once the analytic parton shower is released
as an official part of the \whizard\ package.

\subsection{Handling of beam remnants}
\label{sec:beamremnants}

We implemented a very rudimentary treatment of beam remnants, with the main purpose of being able to provide a color-neutral input to the hadronization. In dependence of the emitted particle the beam remnant is assumed to consist of one or two partons, the procedure for determining these partons' flavours and momenta is given below.

The given procedure obviously only applies in the case of only one emitted parton per proton, that is in the case of only one hard interaction. As an implementation of an interleaved multiple interactions/initial-state radiation evolution along the lines of the \emph{Interleaved Evolution} approach \cite{Sjostrand:2004ef} is in preparation, this simple treatment of beam-remnants will become inapplicable. Thus a more sophisticated treatment will be implemented in the future.

\subsubsection{Flavours}

The flavours of the beam remnant are chosen according to a simplified
version of \pythia's procedure \cite[section 11.1.1]
{PYTHIA64}. Depending on the flavour of the emitted parton the
flavours of the beam remnant are chosen (These rules apply for protons
as the initial hadrons, with obvious substitutions for antiprotons.): 
\begin{itemize}
\item A valence quark of the hadron is assumed to leave behind a
  diquark beam remnant. A $ud$-diquark\footnote{Diquarks are given in
    the notation $qq_S$, where the $q$ are the building quarks and $S$
    is the total spin.} is assumed to be a $ud_1$ in 25\% and a $ud_0$
  in 75\% of the cases, while a $uu$-diquark is always a $uu_1$. 
\item A gluon is assumed to leave behind a colour octet state, that is
  divided into a colour triplet quark and an anti-colour triplet
  antiquark. The division into $u$ + $ud_1$ for $1/6$ of the cases,
  into $u$ + $ud_0$ for $1/2$ and into $d$ + $uu_1$ for $1/3$ of the
  cases. 
\item A sea quark, for example a $s$, leaves behind an $uud\bar{s}$
  state, that is subdivided into a meson and a diquark. The relative
  probabilities are $1/6$ into $u\bar{s}$ + $ud_1$, $1/2$ into
  $u\bar{s}$ + $ud_0$ and $1/3$ into $d\bar{s}$ + $uu_1$. 
\item An antiquark $\bar{q}$ leaves behind a $uudq$ state, that is
  divided into a baryon and a quark. Since mostly the $q\bar{q}$ pair
  comes from an emission of a gluon, the subdivision $uud$ + $q$ is
  not allowed as it would correspond to a color singlet gluon. The
  subdivision is therefore in $2/3$ of the cases into $udq$ + $u$ and
  in $1/3$ of the cases into $uuq$ + $d$. The three quark state $uuq$
  or $udq$ is then replaced by the corresponding baryon of lowest
  spin. 
\end{itemize}

\subsubsection{Momenta}
The total momentum of the beam remnant is given by the remaining
momentum of the hadron after the emitted particle has been removed. In
case the beam remnant consists of only one parton, this parton is
assigned the complete momentum, if the beam remnant consists of a
diquark and a quark the momentum is distributed in equal parts to both
the constituents. If the beam remnant consists of two constituents
with one of them being a meson or baryon, the energy is distributed in
equal parts but the three-momentum is distributed so that the hadron
is on-shell and the quark is assigned the remaining momentum. This
procedure generates on-shell colorless particles and off-shell
coloured particles so that the coloured particles off-shellness is
absorbed in the hadronization. 

\section{Results and Validation}
\label{sec:results}

We compared the predictions for the process $e^+ e^- \rightarrow
hadrons$ at LEP with an energy of $\sqrt{s}=133\ei{GeV}$ and for $Z$
production, $p\bar{p}/pp \rightarrow Z + X$, at the Tevatron and the
LHC at energies of $\sqrt{s}=1.96\ei{TeV}$ and $\sqrt{s}=7\ei{TeV}$, respectively. All event sets were generated using \whizard, which means the hard interaction was simulated by \whizard/\texttt{O'Mega}, the parton shower was either simulated using \pythia's virtuality-ordered shower or \whizard's own analytic shower, denoted in the plots by either \pythia\ PS or \whizard\ PS. 
For \whizard's parton shower we used a first-order running $\alpha_S$ in the $\overline{\mbox{MS}}$-scheme, given by
\begin{equation}
  \alpha_S \left(Q^2\right) = \frac{4 \pi} { \left( 11 -\frac23 n_f \right) \log\left(Q^2 / \Lambda^2 \right) },
\end{equation}
taking $n_f$ and $\Lambda$ as constants, neglecting the influence of flavor thresholds for the moment. For \pythia's parton shower we used the same $\Lambda$ and $n_f$ values for a first-oder running $\alpha_S$, but with threshold effects enabled. If not stated otherwise, the values for $\Lambda$ were chosen to be $\Lambda = 0.19 \ei{GeV}$ for \whizard's parton shower and $\Lambda = 0.29 \ei{GeV}$ for \pythia's parton shower. The $\Lambda$ value for \whizard\ was chosen by hand to improve agreement with \pythia's distributions.
The hadronization, if activated, was simulated using \pythia\ with the
hadronization tune from \cite[table 10, Dec. 93]{DELPHI2}. This tune
was of course made using \pythia's parton shower, but will be used in
here together with \whizard's parton shower as well. As the
hadronization tune depends on the parton shower, using a tuning
obtained with a different parton shower can lead to unsubstantial
deviations in the results. As there is no tune with \whizard's parton
shower available, we cannot give an estimate for the deviations. The
possible tuning of our shower is beyond the scope of the present
paper, presenting merely the algorithm, and will be left for future
work. Given the fact that no tuning has been done, the shower already
describes data in a QCD environment reasonably well.

The definitions of all observables are given in section \ref{sec:definitions} in the appendix.

\subsection{Final State Radiation at parton level}

\begin{figure}
 \centering
 \psfrag{1-T}[l][][1][0] {$1-T$}
 \psfrag{Tmajor}[l][][1][0] {$T_{major}$}
 \psfrag{1/N dN/d(1-T)}[l][][1][0] {$1/N \;\dif N / \dif (1-T)$}
 \psfrag{1/N dN/dTmajor}[l][][1][0] {$1/N \;\dif N / \dif T_{major}$}
 \includegraphics[bb=50 50 554 770,scale=0.45, angle=-90]{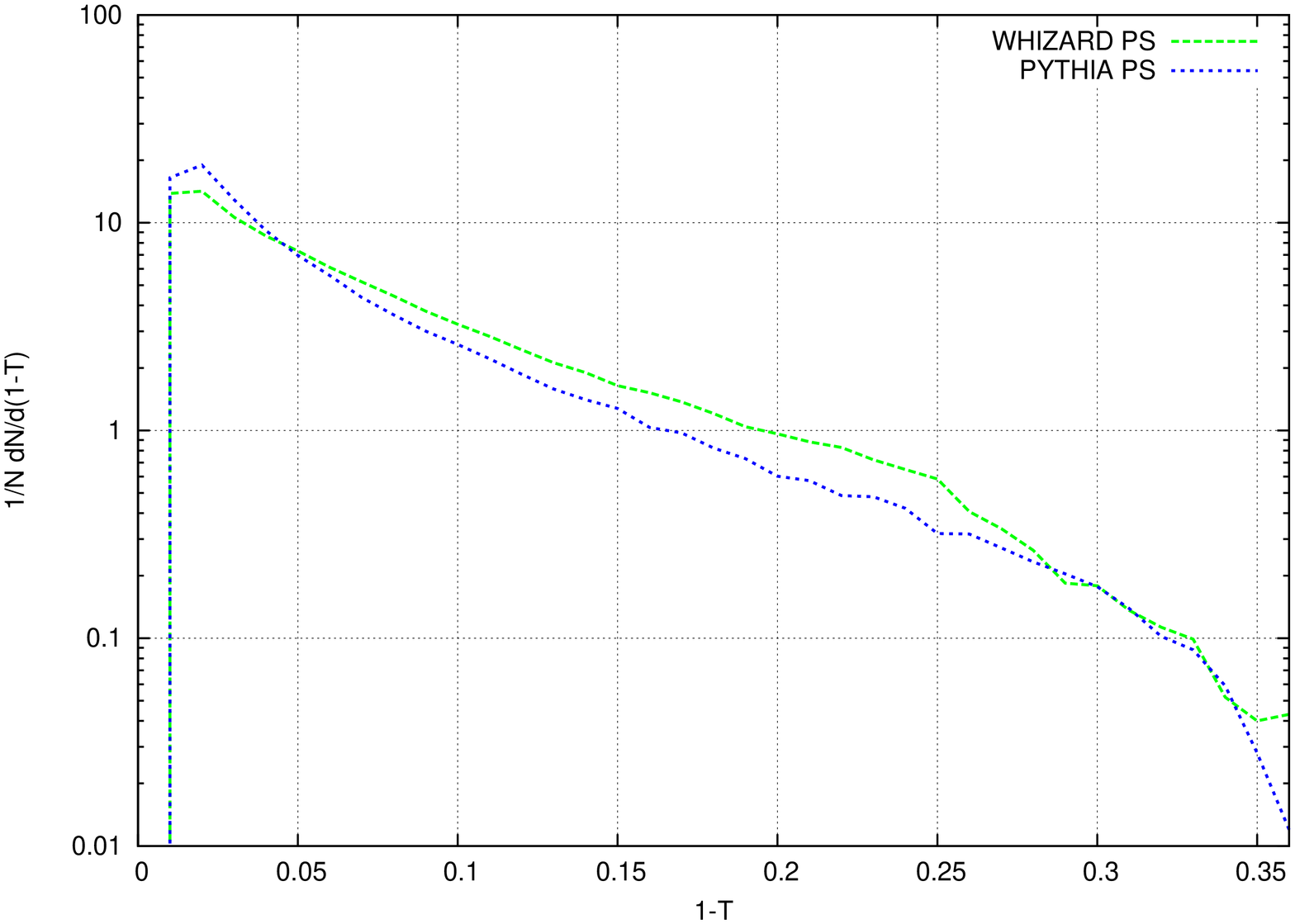}
 \includegraphics[bb=50 50 554 770,scale=0.45, angle=-90]{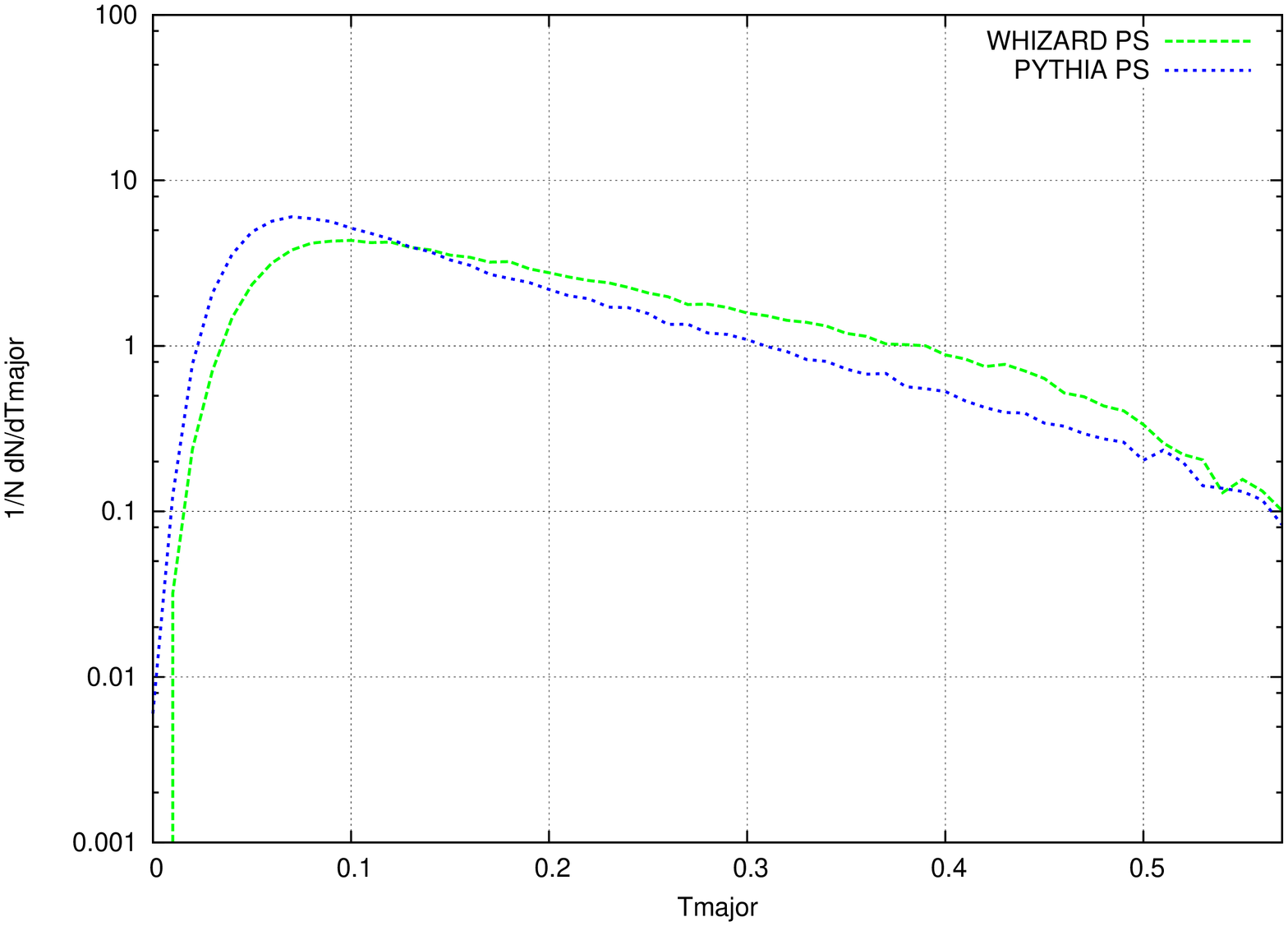}
 \caption{Plots for thrust $T$ and thrust major $T_{major}$ (without hadronization). The dashed/green/bright line is \whizard, the dotted/blue/dark line is \pythia.}
 \label{fig:fsrplotsnohad1}
\end{figure}
\begin{figure}
 \centering
 \psfrag{Tminor}[l][][1][0] {$T_{minor}$}
 \psfrag{Oblateness}[l][][1][0] {$O$}
 \psfrag{1/N dN/dTminor}[l][][1][0] {$1/N \;\dif N / \dif T_{minor}$}
 \psfrag{1/N dN/dO}[l][][1][0] {$1/N \;\dif N / \dif O$}
 \includegraphics[bb=50 50 554 770,scale=0.45, angle=-90]{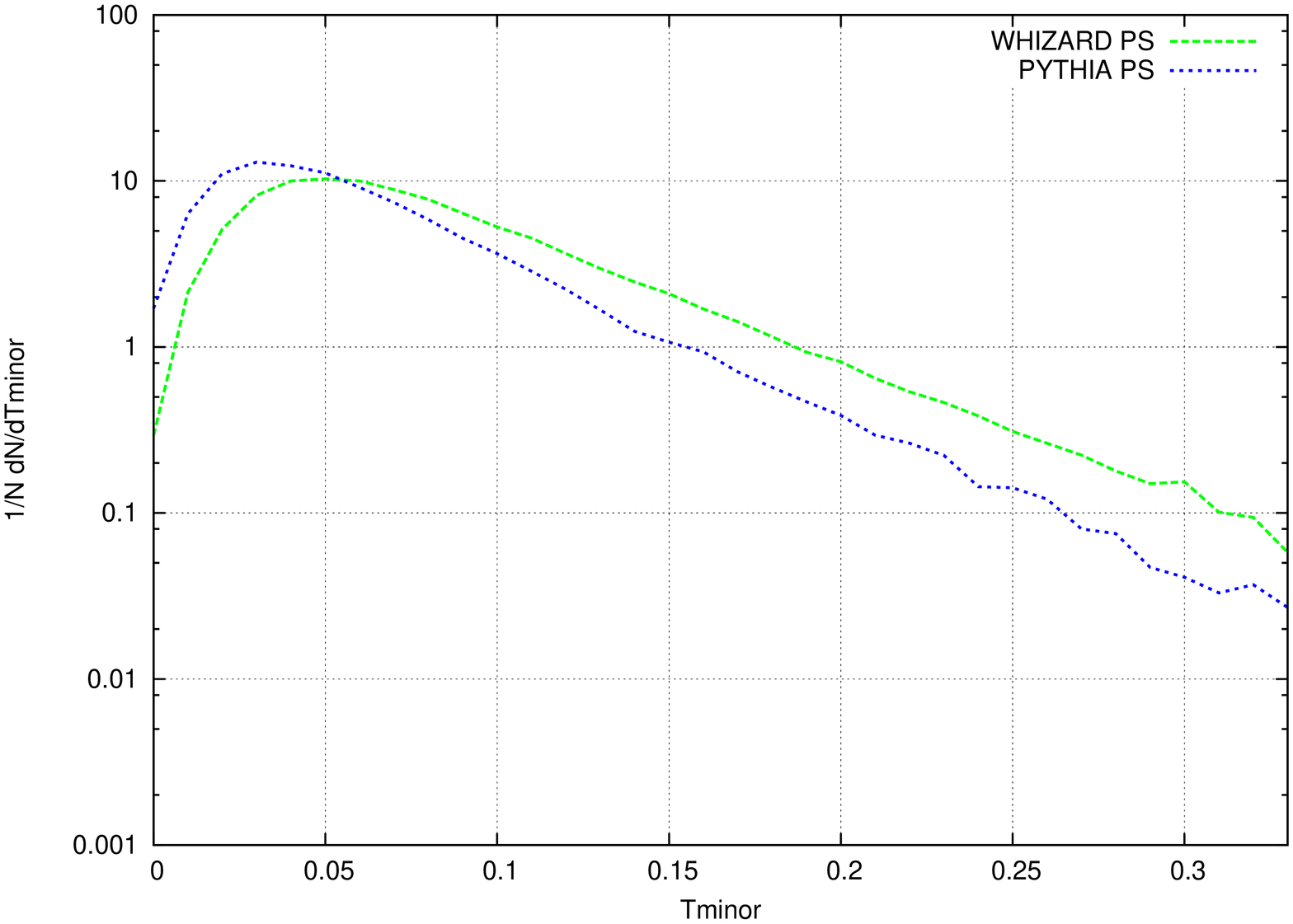}
 \includegraphics[bb=50 50 554 770,scale=0.45, angle=-90]{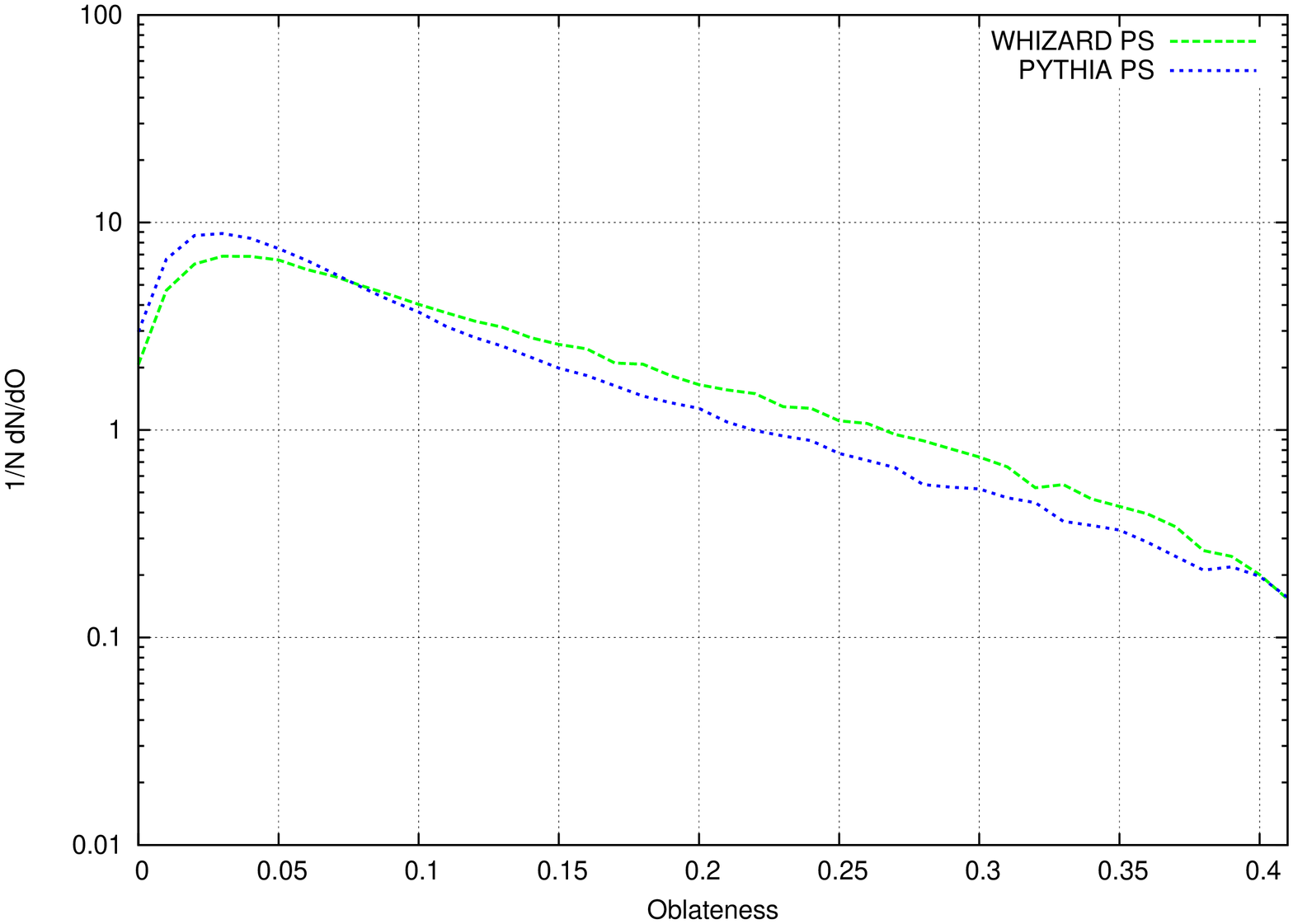}
 \caption{Plots for thrust minor $T_{minor}$ and Oblateness $O$ (without hadronization). The dashed/green/bright line is \whizard, the dotted/blue/dark line is \pythia.}
 \label{fig:fsrplotsnohad2}
\end{figure}
\begin{figure}
 \centering
 \psfrag{Bmax}[l][][1][0] {$B_{max}$}
 \psfrag{Bmin}[l][][1][0] {$B_{min}$}
 \psfrag{1/N dN/dBmax}[l][][1][0] {$1/N \;\dif N / \dif B_{max}$}
 \psfrag{1/N dN/dBmin}[l][][1][0] {$1/N \;\dif N / \dif B_{min}$}
 \includegraphics[bb=50 50 554 770,scale=0.45, angle=-90]{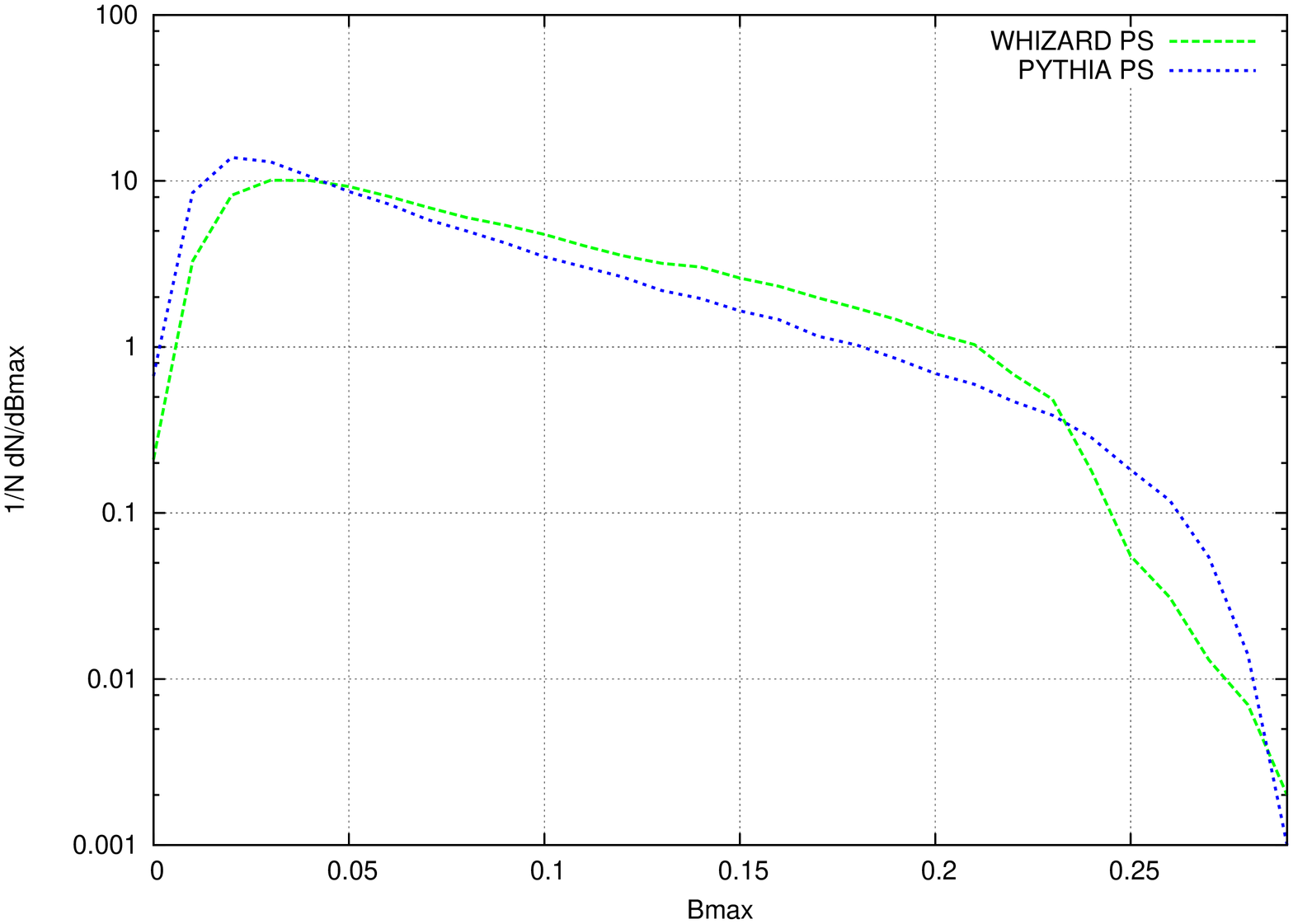}
 \includegraphics[bb=50 50 554 770,scale=0.45, angle=-90]{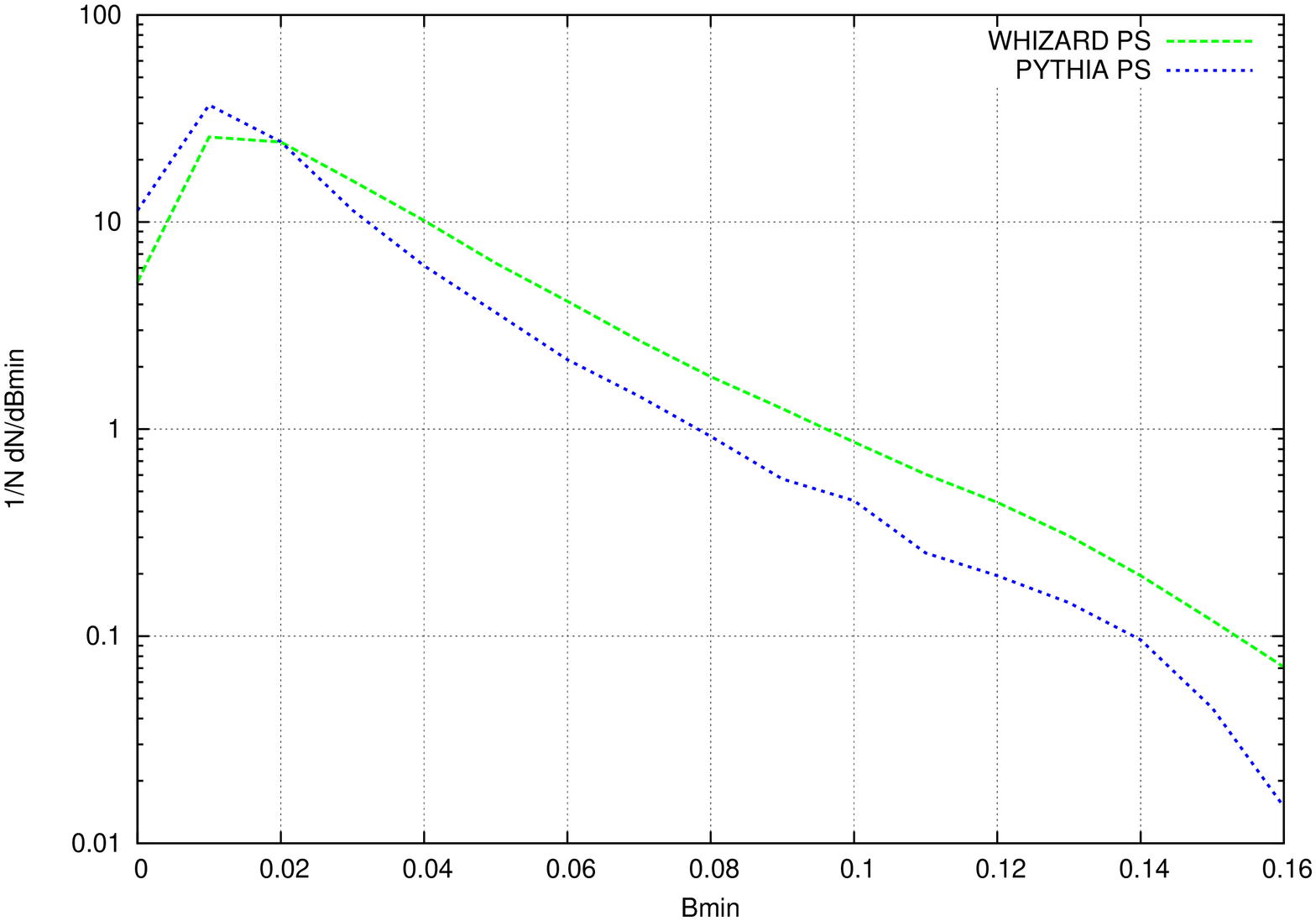}
 \caption{Plots for jet broadenings $B_{max}$ and $B_{min}$ (without hadronization). The dashed/green/bright line is \whizard, the dotted/blue/dark line is \pythia.}
 \label{fig:fsrplotsnohad3}
\end{figure}
\begin{figure}
 \centering
 \psfrag{Bsum}[l][][1][0] {$B_{sum}$}
 \psfrag{Bdiff}[l][][1][0] {$B_{diff}$}
 \psfrag{1/N dN/dBsum}[l][][1][0] {$1/N \;\dif N / \dif B_{sum}$}
 \psfrag{1/N dN/dBdiff}[l][][1][0] {$1/N \;\dif N / \dif B_{diff}$}
 \includegraphics[bb=50 50 554 770,scale=0.45, angle=-90]{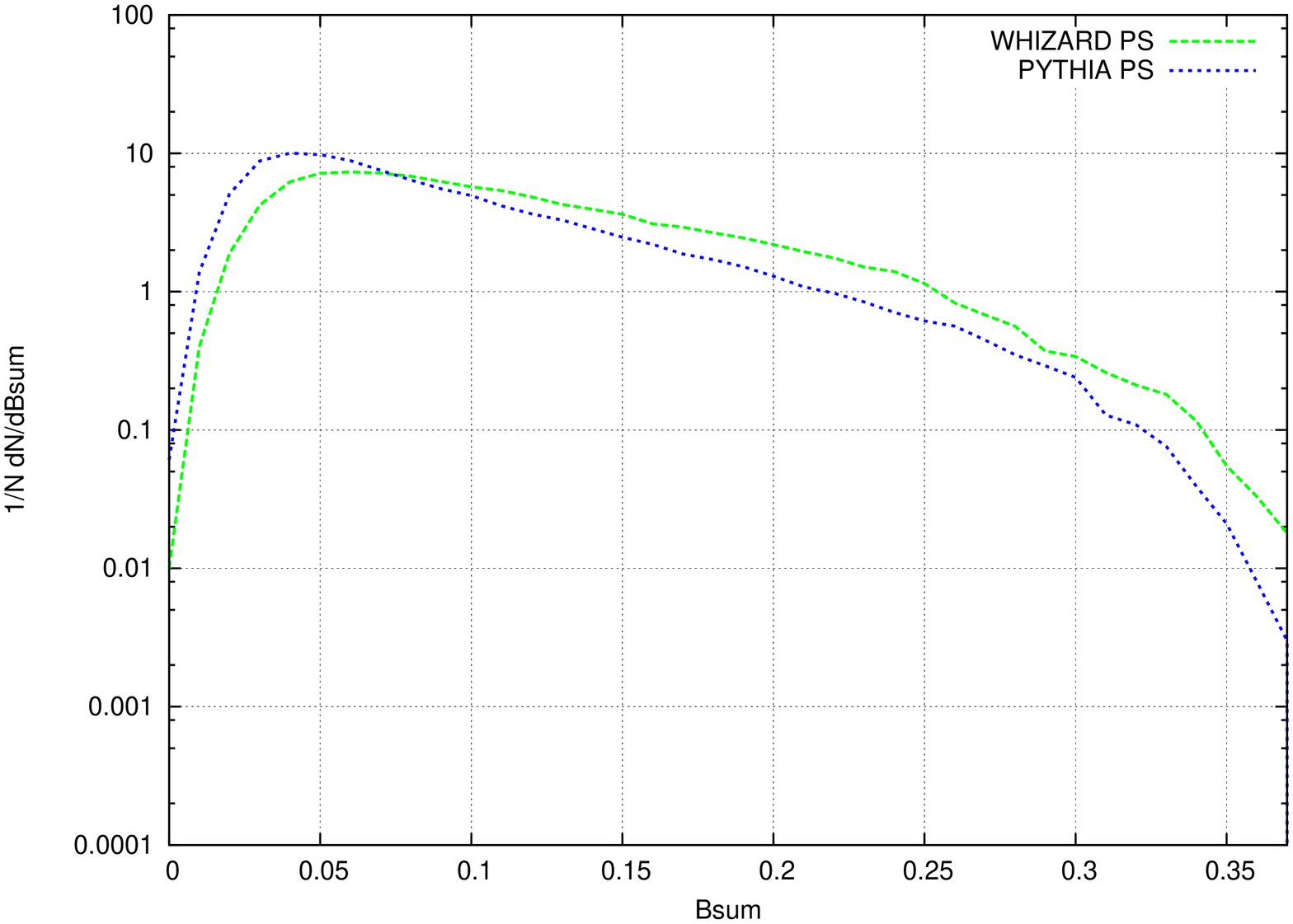}
 \includegraphics[bb=50 50 554 770,scale=0.45, angle=-90]{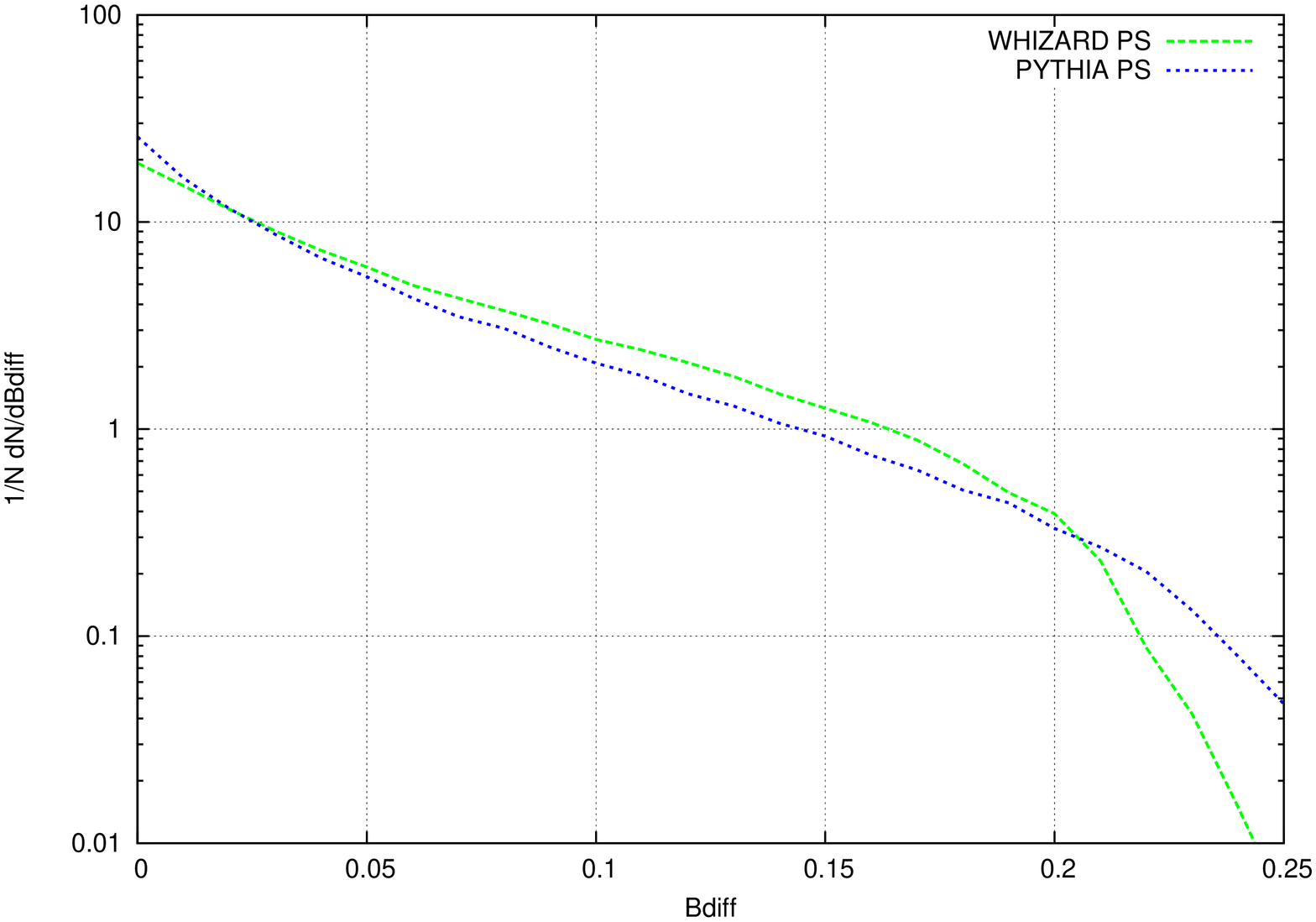}
 \caption{Plots for jet broadenings $B_{sum}$ and $B_{diff}$ (without hadronization). The dashed/green/bright line is \whizard, the dotted/blue/dark line is \pythia.}
 \label{fig:fsrplotsnohad4}
\end{figure}

Figures \ref{fig:fsrplotsnohad1} to \ref{fig:fsrplotsnohad4} show a
comparison of distributions of event shapes at parton level. For both
parton shower programs the events were showered with a cut-off
virtuality $Q^2_{min} = 1 \ei{GeV}$, hadronization was disabled. The
plots for thrust, thrust major and thrust minor show that \whizard's
parton shower generates more spherical events compared to \pythia's
parton shower. Nonetheless, they show a satisfactory agreement as
\whizard's parton shower was not tuned at all for these
plots. However, it is unclear if the discrepancies can be tuned
away. Moreover, as distributions at parton level are not observable in
an experiment, it is doubtful if they need to be. 

\subsection{Final State Radiation at hadron level}
\label{sec:FSRhad}

\subsubsection{Event shapes}

For hadronized events, we can compare the generated distributions with
experimental data. We compared the distributions for several event
shapes with data from the DELPHI collaboration \cite{DELPHI1}. The
measurement was performed using $e^+ e^-$ collisions at center-of-mass
energies of $\sqrt{s} = 130\ei{GeV}$ and $136\ei{GeV}$. The simulated
hard interaction was chosen to be $e^+e^- \rightarrow u\bar{u}$ at a
center-of-mass energy of $\sqrt{s} = 133\ei{GeV}$.  

\begin{figure}
 \centering
 \psfrag{1-T}[l][][1][0] {$1-T$}
 \psfrag{Tmajor}[l][][1][0] {$T_{major}$}
 \psfrag{1/N dN/d(1-T)}[l][][1][0] {$1/N \;\dif N / \dif (1-T)$}
 \psfrag{1/N dN/dTmajor}[l][][1][0] {$1/N \;\dif N / \dif T_{major}$}
 \includegraphics[bb=50 50 554 770,scale=0.45, angle=-90]{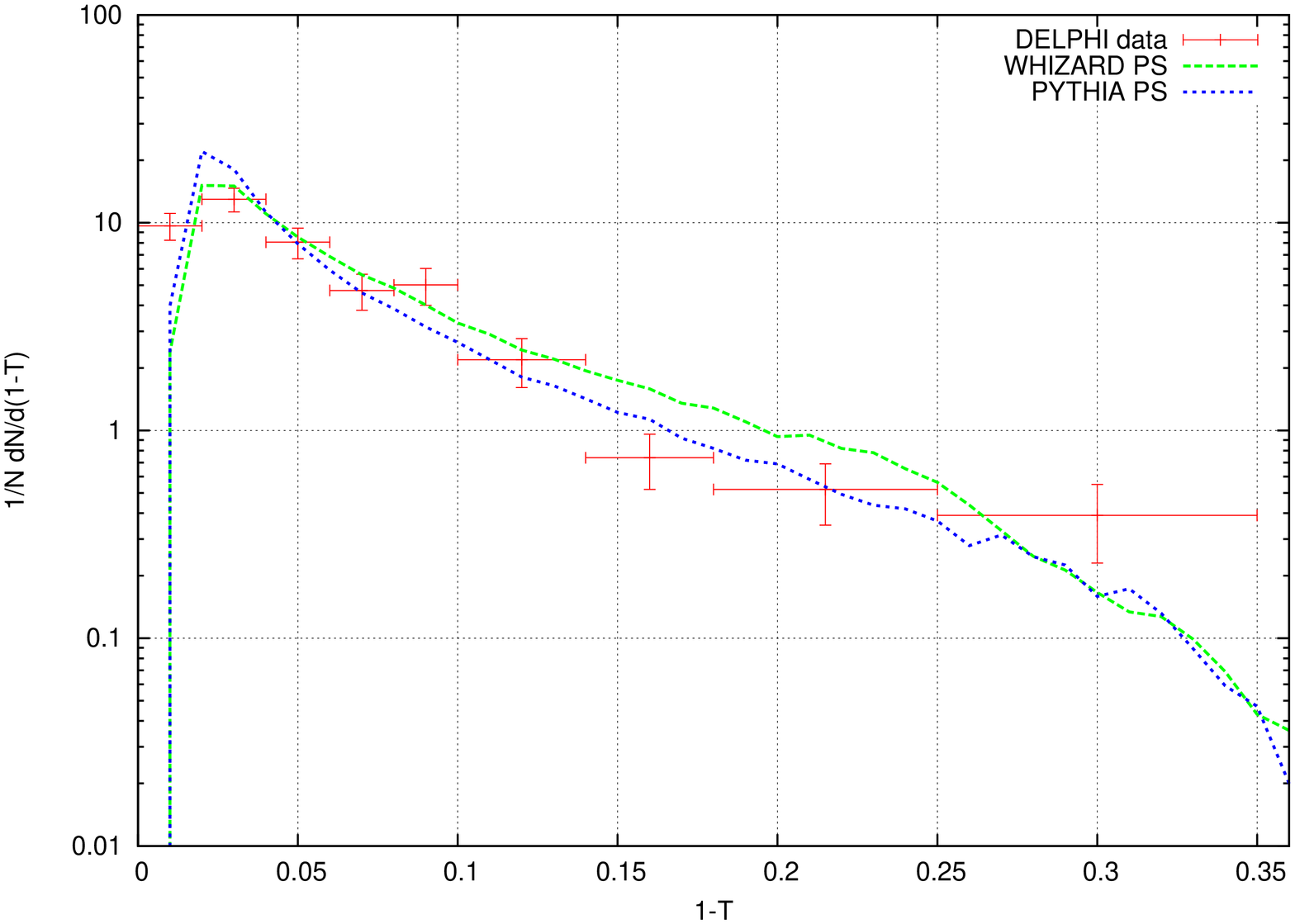}
 \includegraphics[bb=50 50 554 770,scale=0.45, angle=-90]{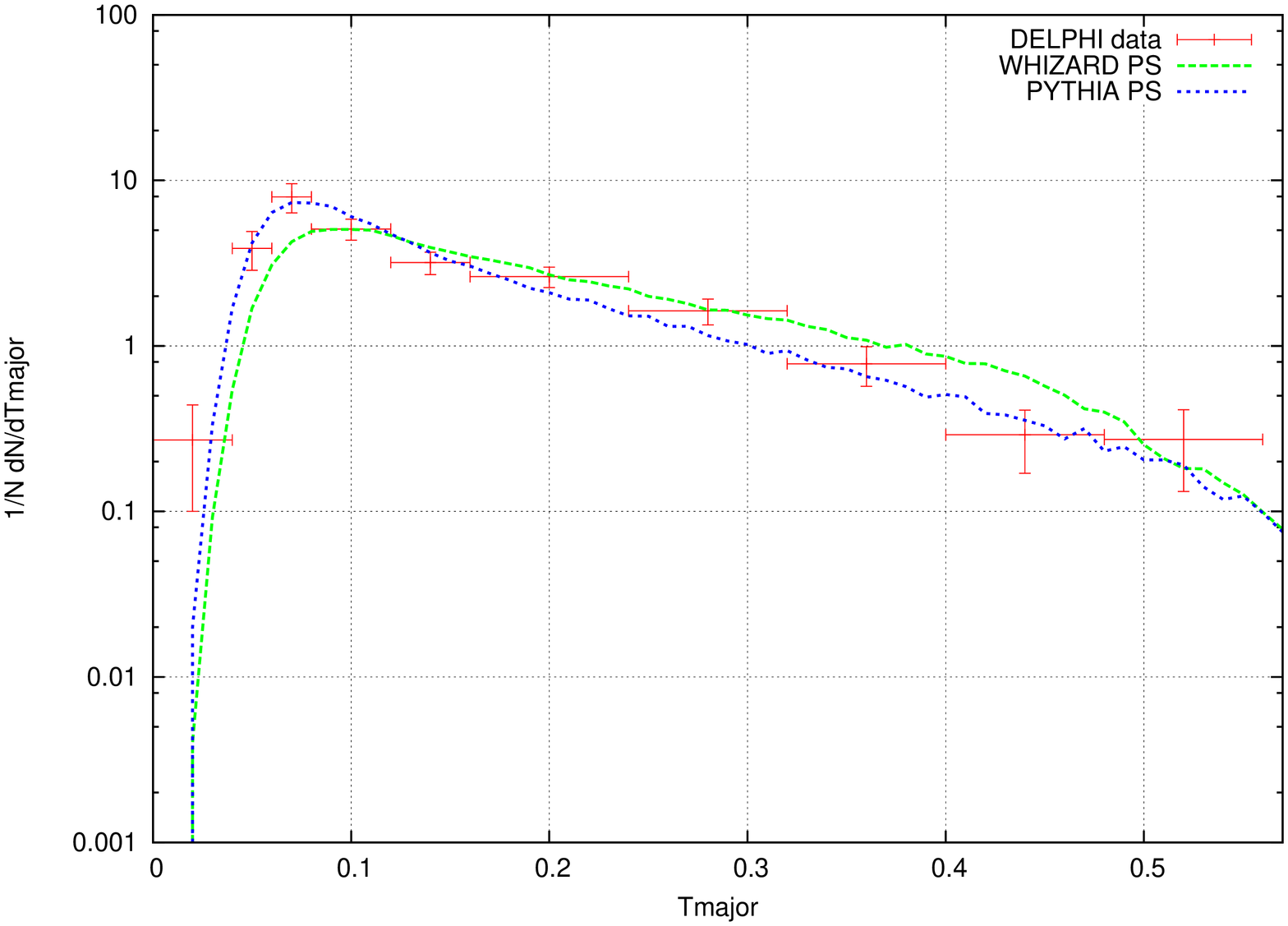}
 \caption{Plots for thrust $T$ and thrust major $T_{major}$ (with hadronization, data from \cite{DELPHI1}). The dashed/green/bright line is \whizard, the dotted/blue/dark line is \pythia.}
 \label{fig:fsrplotshad1}
\end{figure}

\begin{figure}
 \centering
 \psfrag{Tminor}[l][][1][0] {$T_{minor}$}
 \psfrag{Oblateness}[l][][1][0] {$O$}
 \psfrag{1/N dN/dTminor}[l][][1][0] {$1/N \;\dif N / \dif T_{minor}$}
 \psfrag{1/N dN/dO}[l][][1][0] {$1/N \;\dif N / \dif O$}
 \includegraphics[bb=50 50 554 770,scale=0.45, angle=-90]{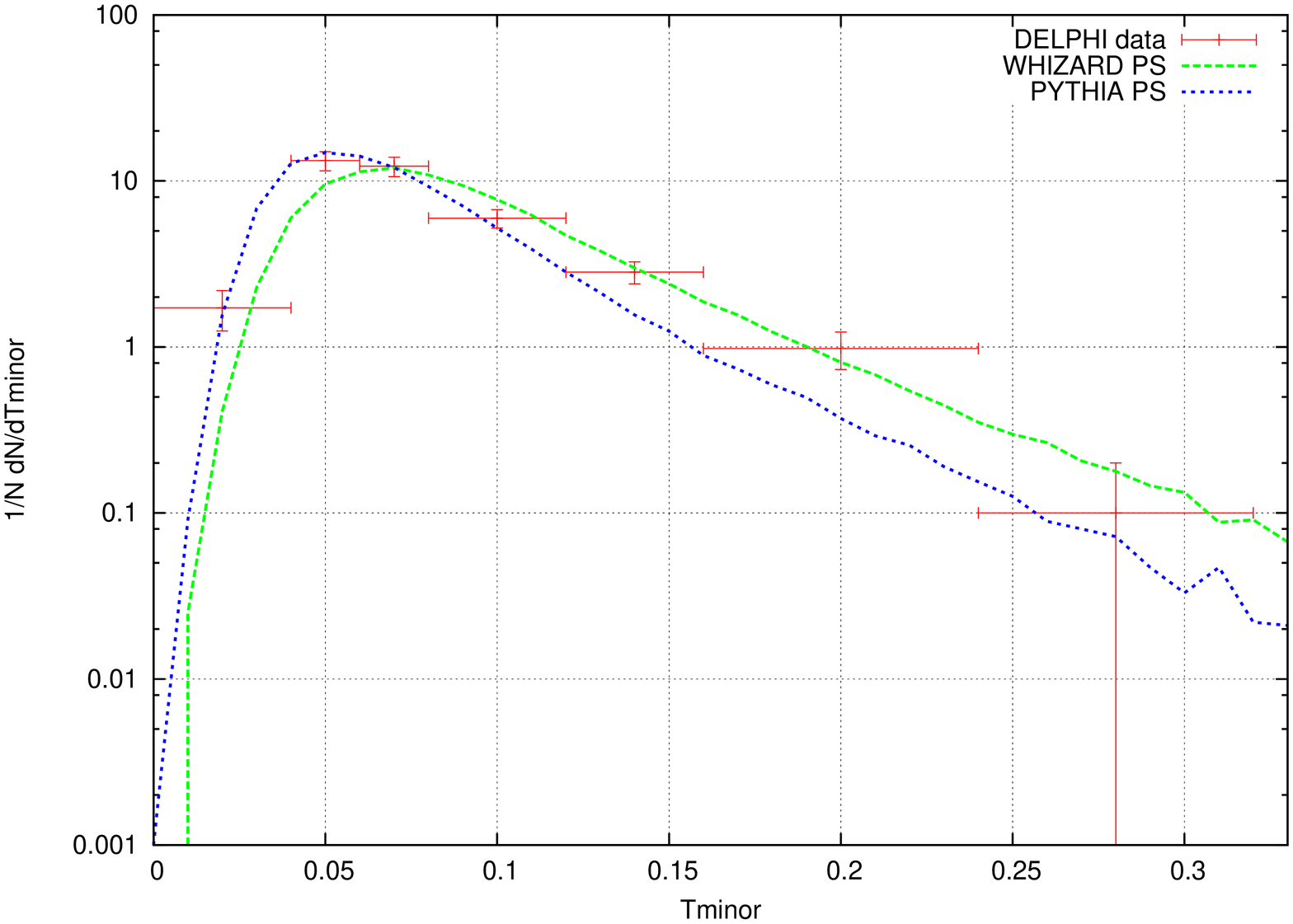}
 \includegraphics[bb=50 50 554 770,scale=0.45, angle=-90]{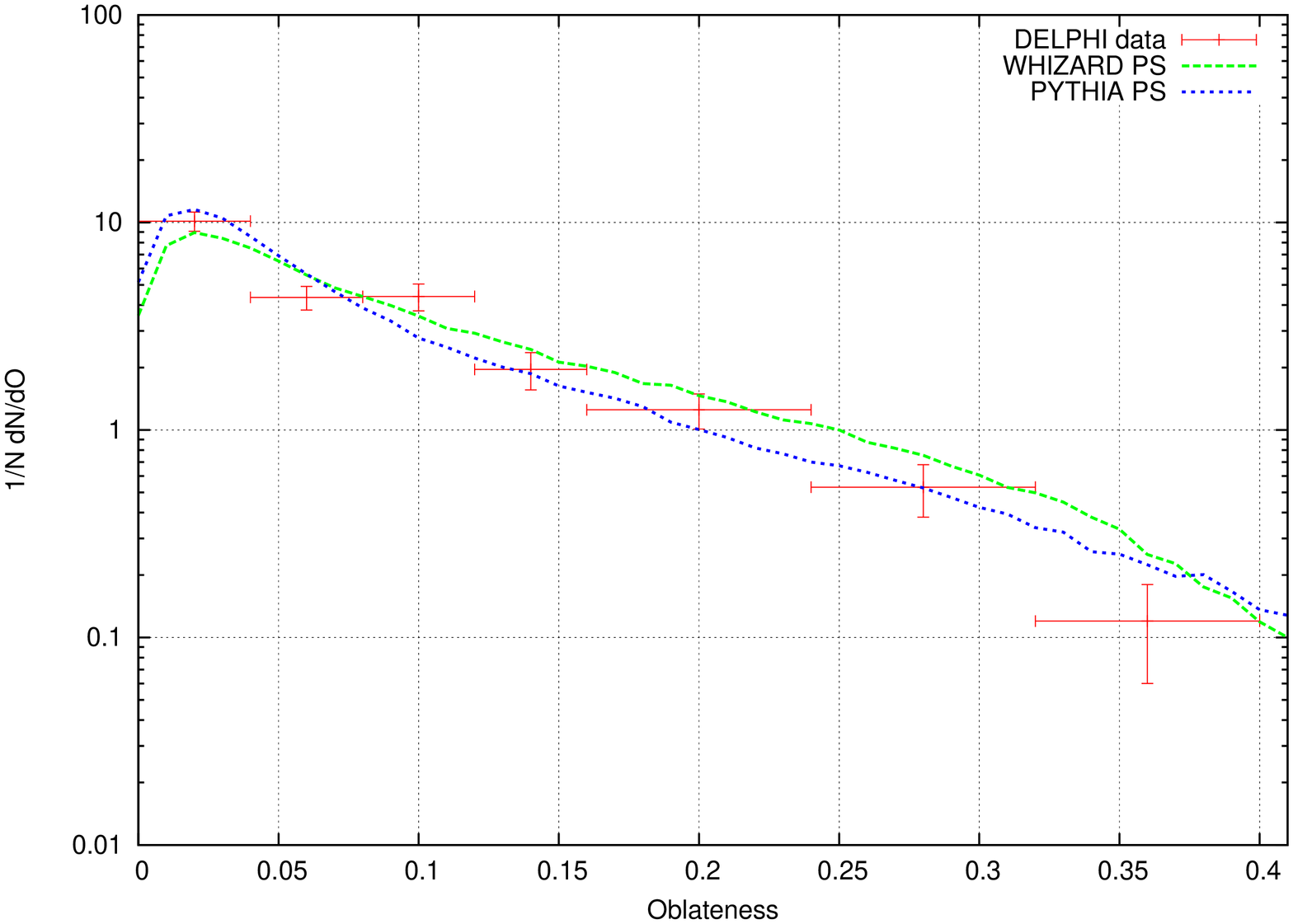}
 \caption{Plots for thrust minor $T_{minor}$ and Oblateness $O$ (with hadronization, data from \cite{DELPHI1}). The dashed/green/bright line is \whizard, the dotted/blue/dark line is \pythia.}
 \label{fig:fsrplotshad2}
\end{figure}

\begin{figure}
 \centering
 \psfrag{Bmax}[l][][1][0] {$B_{max}$}
 \psfrag{Bmin}[l][][1][0] {$B_{min}$}
 \psfrag{1/N dN/dBmax}[l][][1][0] {$1/N \;\dif N / \dif B_{max}$}
 \psfrag{1/N dN/dBmin}[l][][1][0] {$1/N \;\dif N / \dif B_{min}$}
 \includegraphics[bb=50 50 554 770,scale=0.45, angle=-90]{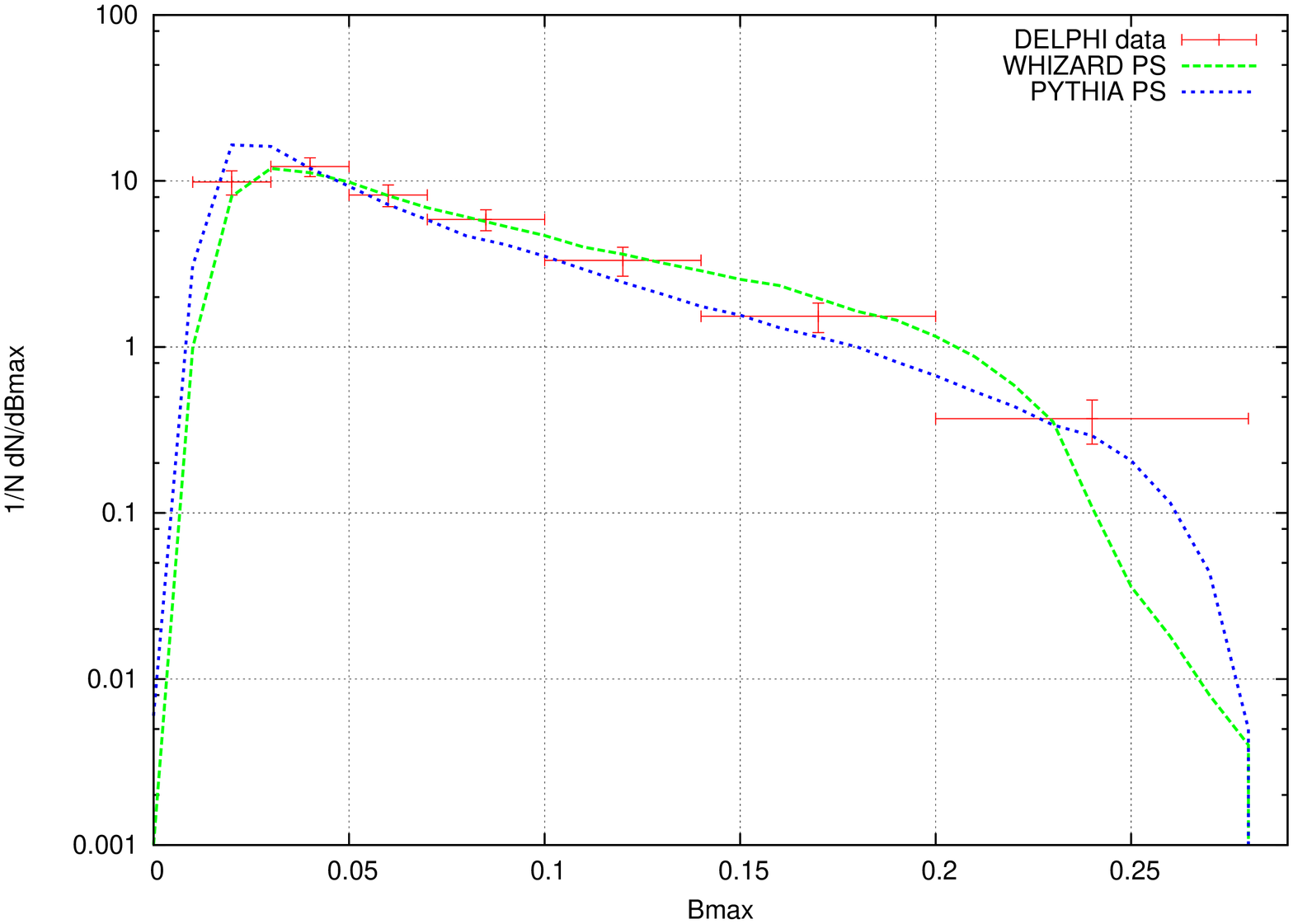}
 \includegraphics[bb=50 50 554 770,scale=0.45, angle=-90]{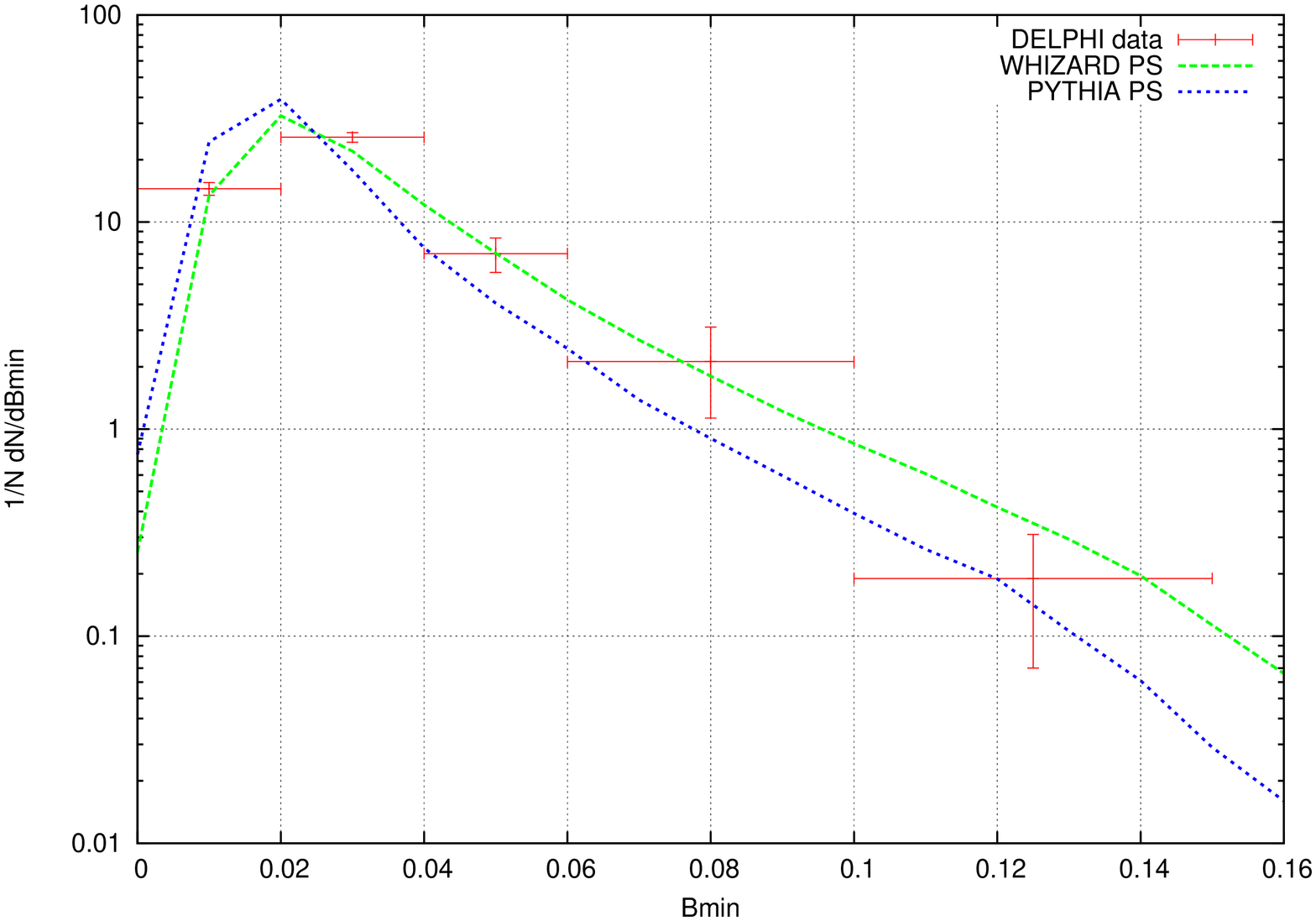}
 \caption{Plots for jet broadenings $B_{max}$ and $B_{min}$ (with hadronization, data from \cite{DELPHI1}). The dashed/green/bright line is \whizard, the dotted/blue/dark line is \pythia.}
 \label{fig:fsrplotshad3}
\end{figure}

\begin{figure}
 \centering
 \psfrag{Bsum}[l][][1][0] {$B_{sum}$}
 \psfrag{Bdiff}[l][][1][0] {$B_{diff}$}
 \psfrag{1/N dN/dBsum}[l][][1][0] {$1/N \;\dif N / \dif B_{sum}$}
 \psfrag{1/N dN/dBdiff}[l][][1][0] {$1/N \;\dif N / \dif B_{diff}$}
 \includegraphics[bb=50 50 554 770,scale=0.45, angle=-90]{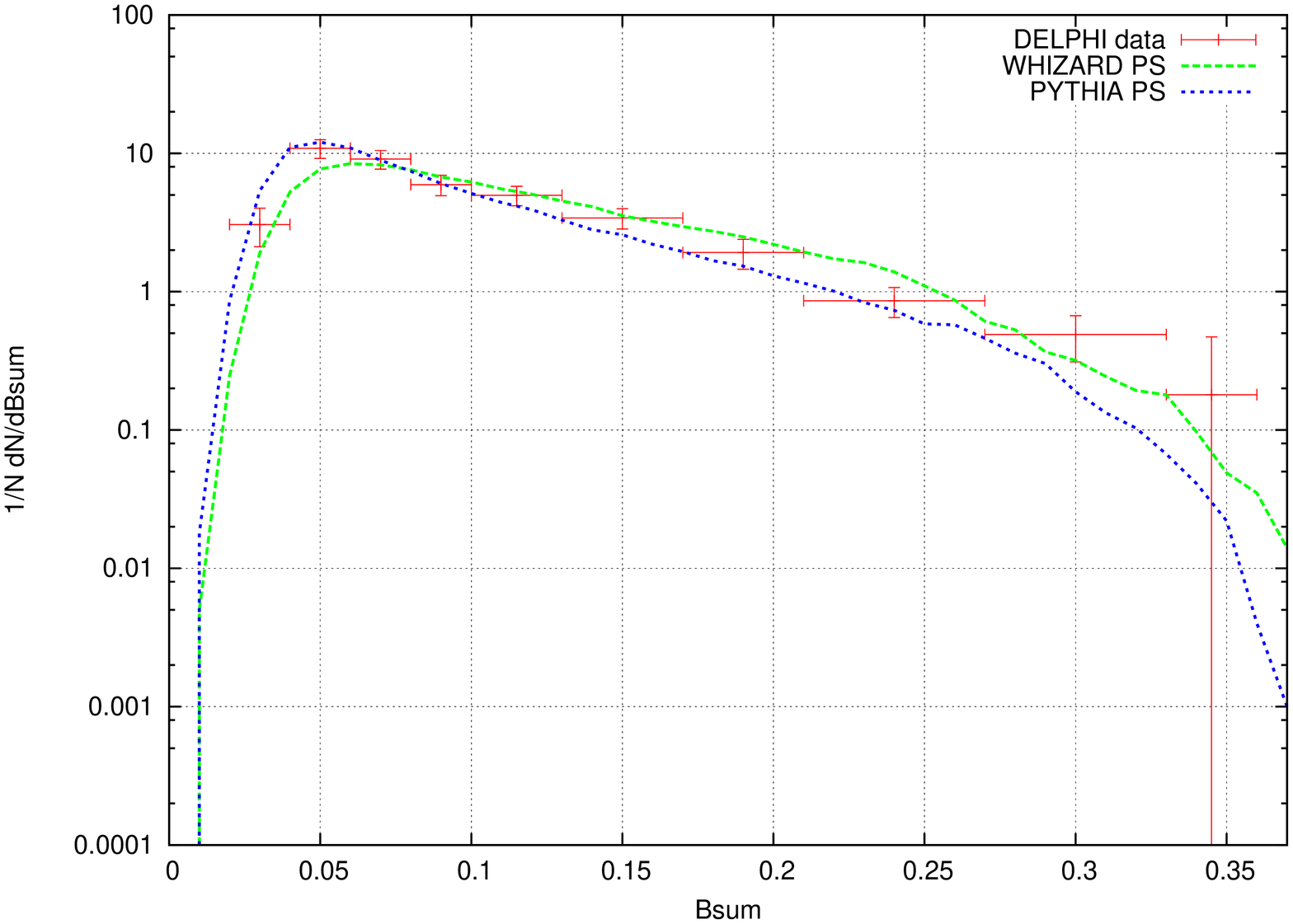}
 \includegraphics[bb=50 50 554 770,scale=0.45, angle=-90]{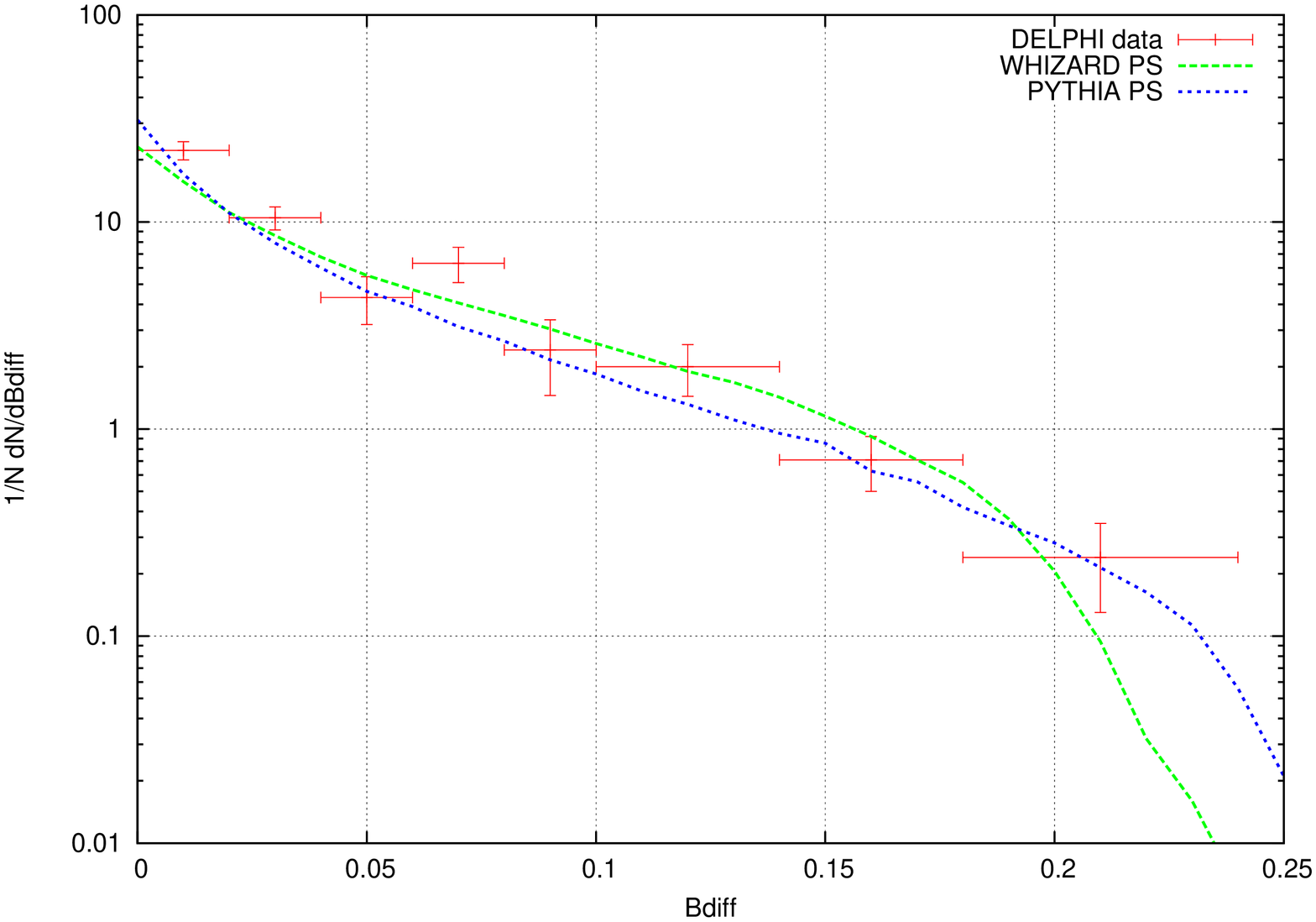}
 \caption{Plots for jet broadenings $B_{sum}$ and $B_{diff}$ (with hadronization, data from \cite{DELPHI1}). The dashed/green/bright line is \whizard, the dotted/blue/dark line is \pythia.}
 \label{fig:fsrplotshad4}
\end{figure}

The results are shown in figures \ref{fig:fsrplotshad1} to
\ref{fig:fsrplotshad4}. 
Both parton showers show good agreement, especially if one takes into
account that the events showered with \whizard's parton shower where
hadronized with the \pythia\ hadronization tuned to data using events
showered with \pythia. As for the unhadronized samples, events
showered with \whizard\ tend to populate the regions corresponding to
more spherical configurations compared to events generated using the
\pythia\ shower. The plot for thrust major $T_{maj}$ shows a slight
undershooting of the \whizard\ curve with respect to the data in the
two bins from 0.04 to 0.08. However, both distributions are mostly
consistent with the data. 


\subsubsection{Jet rates}
\begin{figure}
\begin{center}
\psfrag{y23}[l][][1][0] {\tiny $y_{23}$}
\psfrag{1/N dN/dy23}[l][][1][0] {\tiny $1/N \;\dif N / \dif y_{23}$}
\psfrag{y34}[l][][1][0] {\tiny $y_{34}$}
\psfrag{1/N dN/dy34}[l][][1][0] {\tiny $1/N \;\dif N / \dif y_{34}$}
\psfrag{y45}[l][][1][0] {\tiny $y_{45}$}
\psfrag{1/N dN/dy45}[l][][1][0] {\tiny $1/N \;\dif N / \dif y_{45}$}
\psfrag{y56}[l][][1][0] {\tiny $y_{56}$}
\psfrag{1/N dN/dy56}[l][][1][0] {\tiny $1/N \;\dif N / \dif y_{56}$}
 \includegraphics[bb=50 50 554 770, scale=0.45, angle=-90]{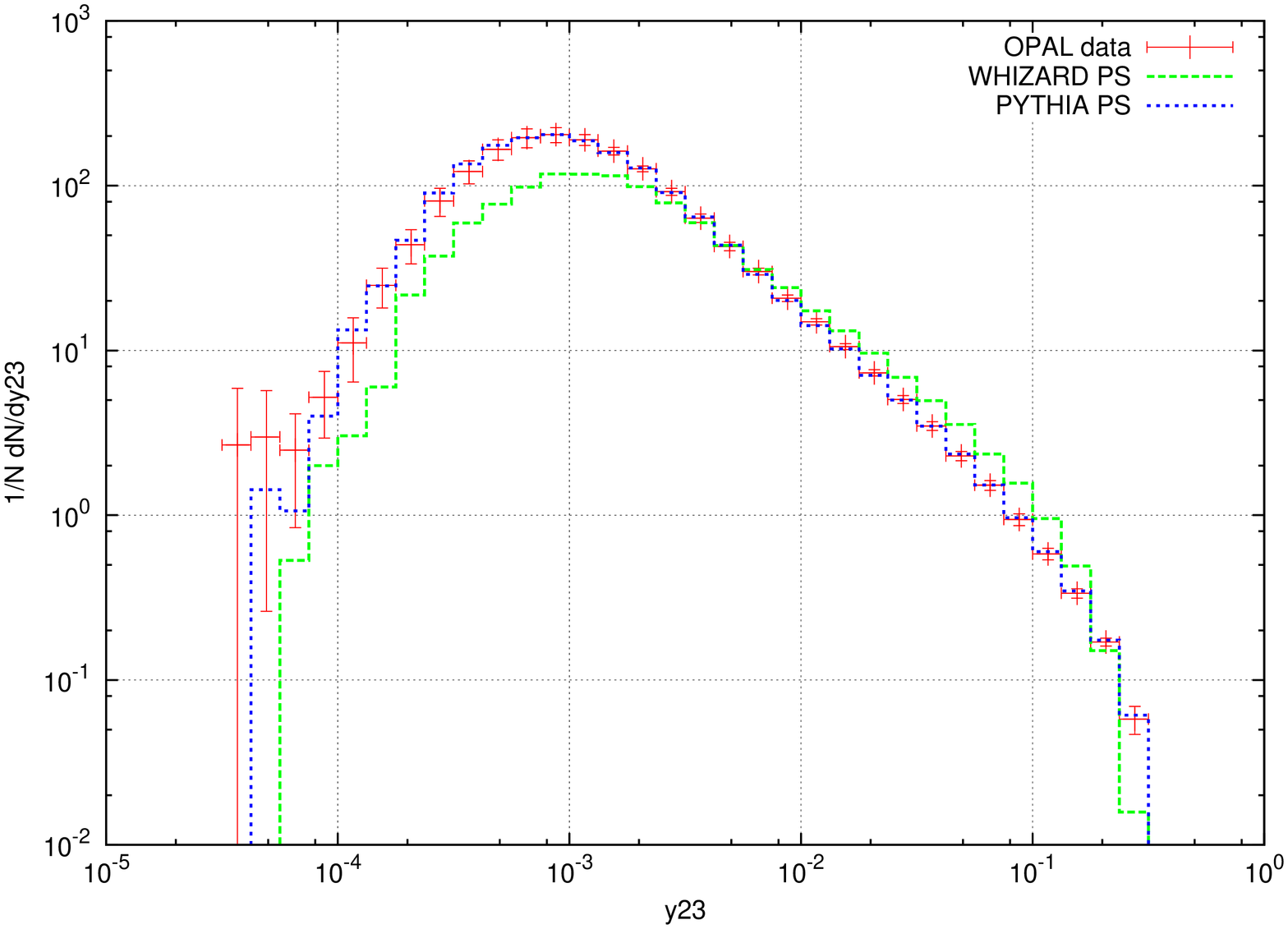}
 \includegraphics[bb=50 50 554 770, scale=0.45, angle=-90]{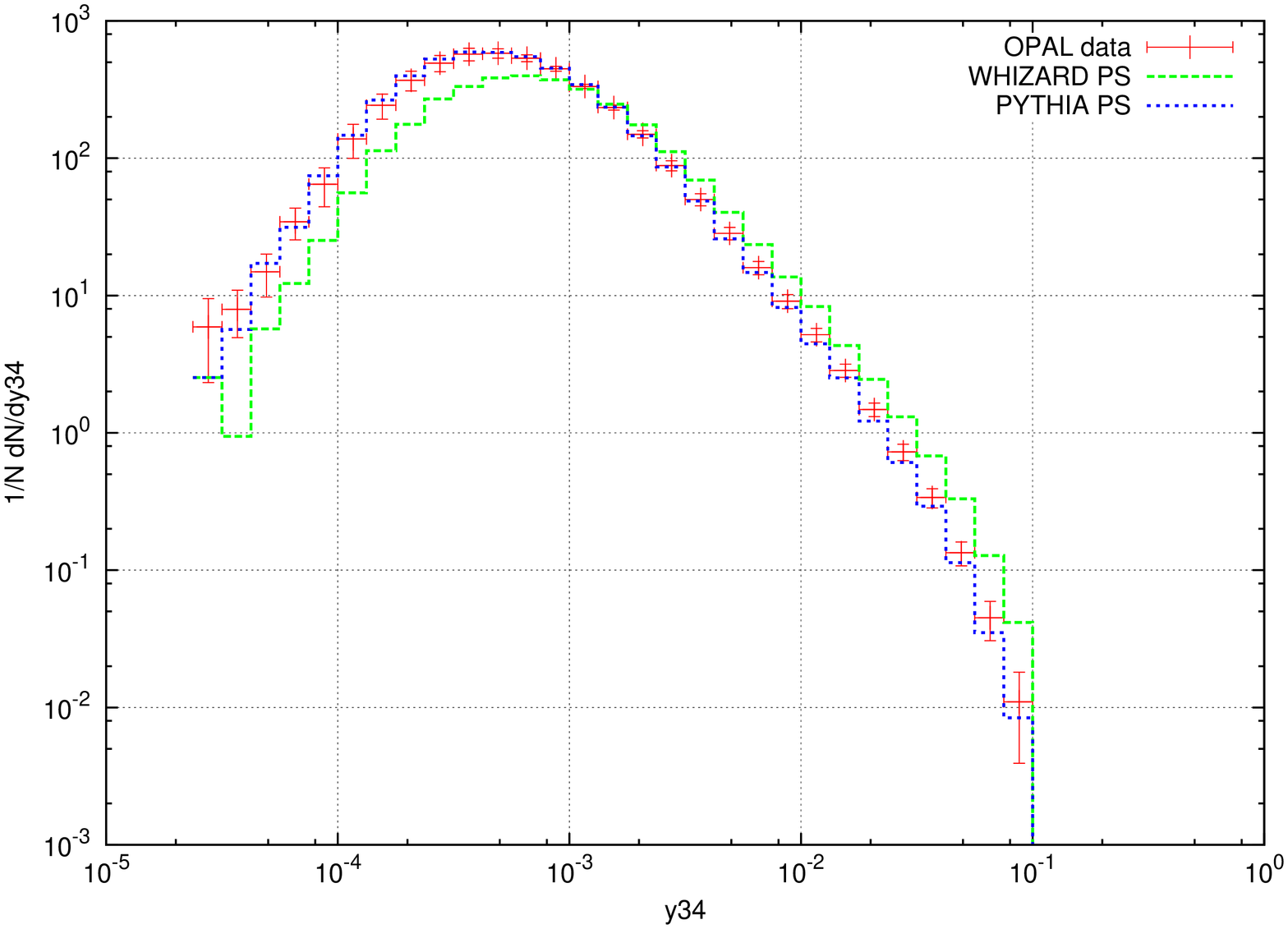}
\caption{Plots for differential jet rates $y_{23}$ and $y_{34}$. The dashed/green/bright line is \whizard, the dotted/blue/dark line is \pythia.}
 \label{fig:fsrYtoHist1}
\end{center}
\end{figure}
\begin{figure}
\begin{center}
\psfrag{y23}[l][][1][0] {\tiny $y_{23}$}
\psfrag{1/N dN/dy23}[l][][1][0] {\tiny $1/N \;\dif N / \dif y_{23}$}
\psfrag{y34}[l][][1][0] {\tiny $y_{34}$}
\psfrag{1/N dN/dy34}[l][][1][0] {\tiny $1/N \;\dif N / \dif y_{34}$}
\psfrag{y45}[l][][1][0] {\tiny $y_{45}$}
\psfrag{1/N dN/dy45}[l][][1][0] {\tiny $1/N \;\dif N / \dif y_{45}$}
\psfrag{y56}[l][][1][0] {\tiny $y_{56}$}
\psfrag{1/N dN/dy56}[l][][1][0] {\tiny $1/N \;\dif N / \dif y_{56}$}
 \includegraphics[bb=50 50 554 770, scale=0.45, angle=-90]{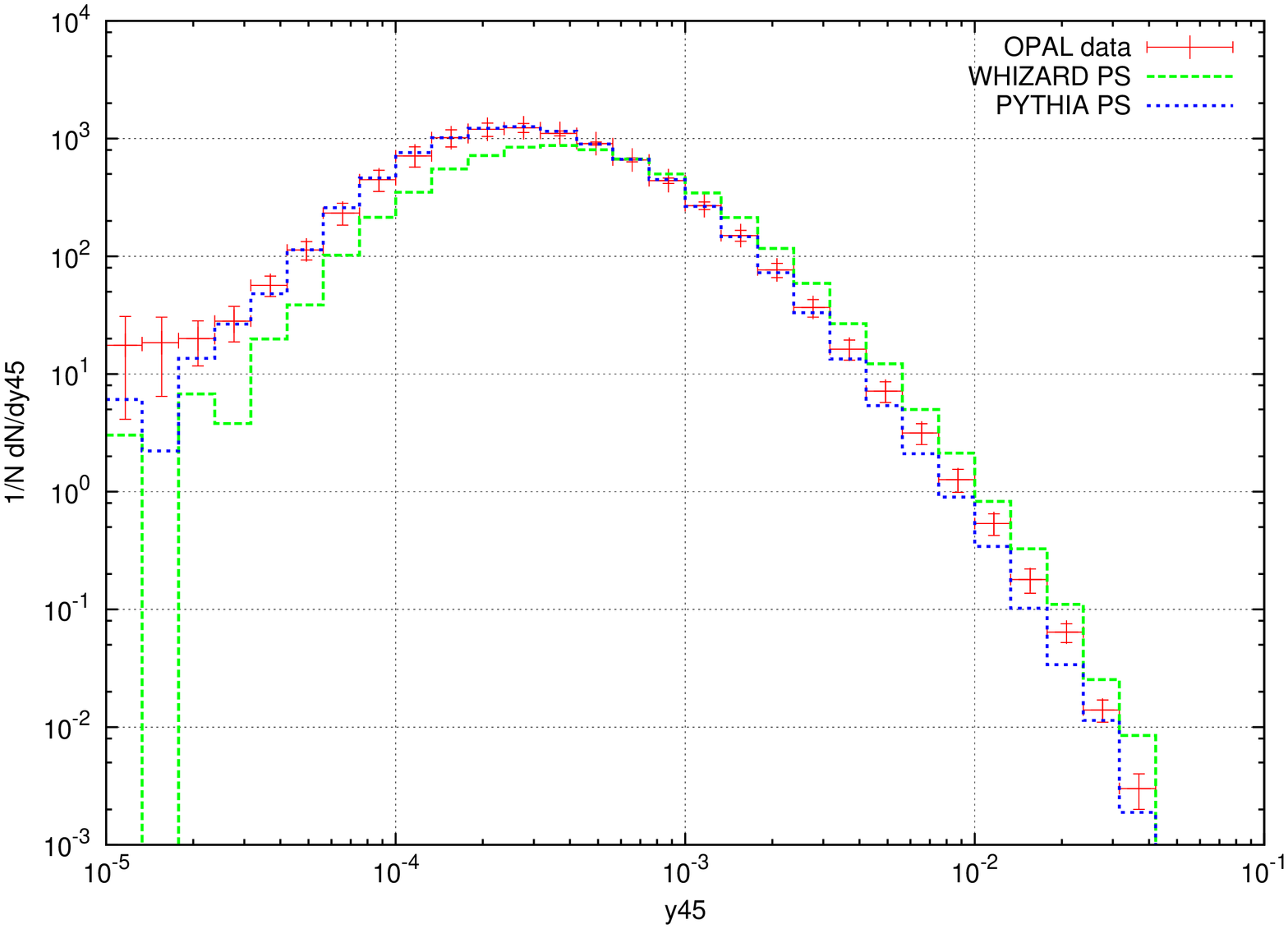}
 \includegraphics[bb=50 50 554 770, scale=0.45, angle=-90]{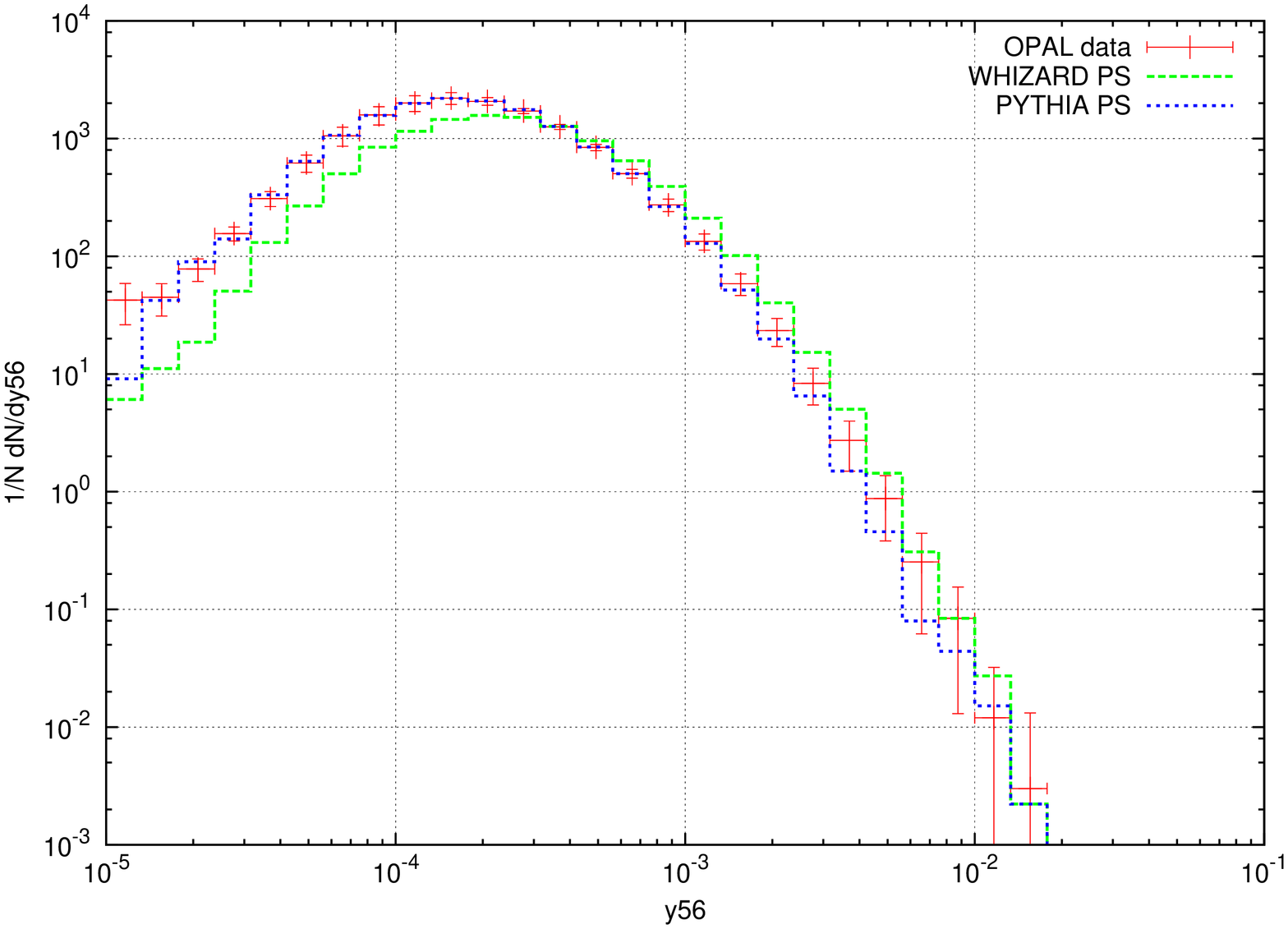}
\caption{Plots for differential jet rates $y_{45}$ and $y_{56}$. The dashed/green/bright line is \whizard, the dotted/blue/dark line is \pythia.}
 \label{fig:fsrYtoHist2}
\end{center}
\end{figure}

A comparison of the Monte Carlo results for the process $e^+ e^- \Rightarrow q\bar{q}$ at $\sqrt{s}=91\ei{GeV}$ with measurements from the JADE and OPAL collaborations given in \cite{OPAL} is shown in figures \ref{fig:fsrYtoHist1} and \ref{fig:fsrYtoHist2}. Shown are differential jet rates as a function of the resolution parameter in the $k_T$-clustering algorithm $y_{i\, i+1}$, where the event turns from being a $i+1$-jet event into a $i$-jet event. The definition of the clustering variable is given in equation (\ref{eq:ktmeasure}) in the appendix. The comparison is equivalent to the one in \cite{Platzer:2011bc,DoktorarbeitSimon}, where a tuning of some parton shower and hadronization parameters was performed. The only tuning applied to the parton shower in the comparison was a by-hand adjustment of $\alpha_S$, setting $\Lambda_{QCD}$ to a value of $0.15\ei{GeV}$. In general, the plots confirm the tendency of \whizard's parton shower to generate more spherical events compared to \pythia\ as the bins with higher values of $y_{i\,i+1}$ are populated more.
Note that small values of $y_{i\,i+1}$ correspond to small invariant masses and that these regions are described by the hadronization model and not the parton shower. So the differences in the left parts of the plots can stem from two sources. They might be caused by normalization effects due to over-estimation in the right parts. Any remaining difference would show that the hadronization tune obtained with \pythia\ is not suitable to describe these regions when used with \whizard's shower.

\subsection{Initial State Radiation}

\begin{figure}
 \centering
 \psfrag{pt}[l][][1][0] {$p_T / \mbox{GeV}$}
 \psfrag{1/N dN/dpt}[l][][1][0] {$1/N \;\dif N / \dif (p_T/\mbox{GeV})$}
 \includegraphics[bb=50 50 554 770,scale=0.45, angle=-90]{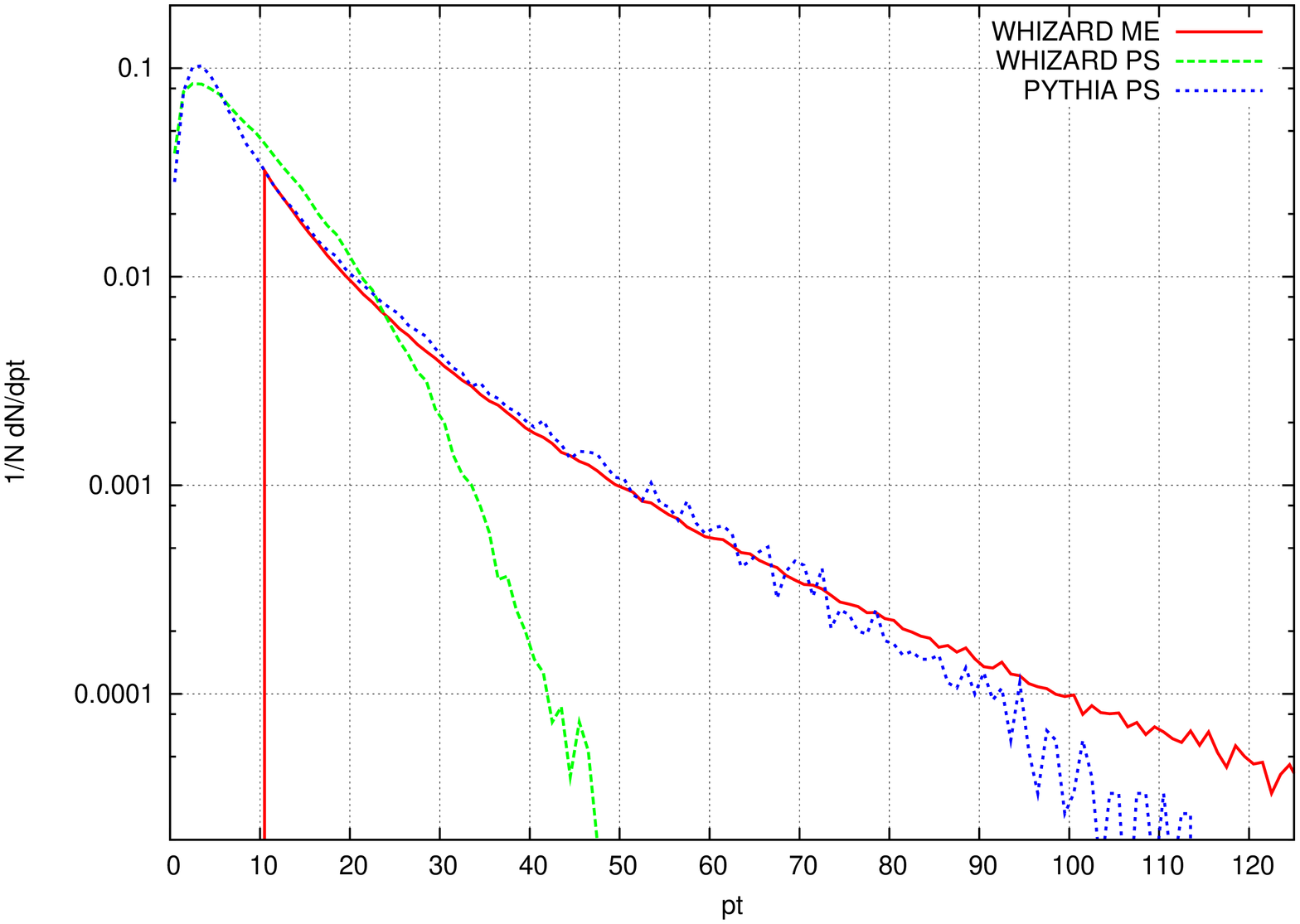}
 \caption{Transverse momentum of a $Z$-Boson in various schemes. The normalization for events from \whizard's matrix element was chosen manually to fit \pythia's PS result in the range $10 \ei{GeV} < p_T < 20 \ei{GeV}$.}
 \label{fig:isr_plot}
\end{figure}

A plot for the transverse momentum of a $Z$-Boson produced in $p\bar{p}$-collisions at $\sqrt{s} = 1.96\ei{TeV}$ is given in figure \ref{fig:isr_plot}. 
The simulation with \pythia\ was done using Rick Field's CDF Tune D6 with CTEQ6L1 parton distribution functions. The simulation using \whizard's parton shower was done using the same PDFs, multiple interactions were disregarded in both simulations.
The data obtained from \whizard's initial-state parton shower shows two distinct features: first of all, the curve in the low-$p_T$ region shows a slight deviation with respect to the corresponding \pythia\ curve. However, as we will see later, this is still in agreement with data. 
Second of all, it shows the known phase space cut at $p_T \lesssim m_Z$ \cite{Miu:1998ju}. For comparison, the plot is supplemented by a $p_T$-histogram for the unshowered process $u\bar{u} \rightarrow Zg$. \pythia's description uses the power-shower and matching and closely resembles the result for the partonic process.

Our approach to solve the shortcomings of \whizard's parton shower was not to include the power shower ansatz, but instead accept this as a deficiency of the parton shower and delegate the task of describing the high-$p_T$ region to a matching algorithm.

\subsection{Matched Final State Radiation}

\begin{figure}
 \centering
 \psfrag{1-T}[l][][1][0] {$1-T$}
 \psfrag{Tmajor}[l][][1][0] {$T_{major}$}
 \psfrag{1/N dN/d(1-T)}[l][][1][0] {$1/N \;\dif N / \dif (1-T)$}
 \psfrag{1/N dN/dTmajor}[l][][1][0] {$1/N \;\dif N / \dif T_{major}$}
 \includegraphics[bb=50 50 554 770,scale=0.45, angle=-90]{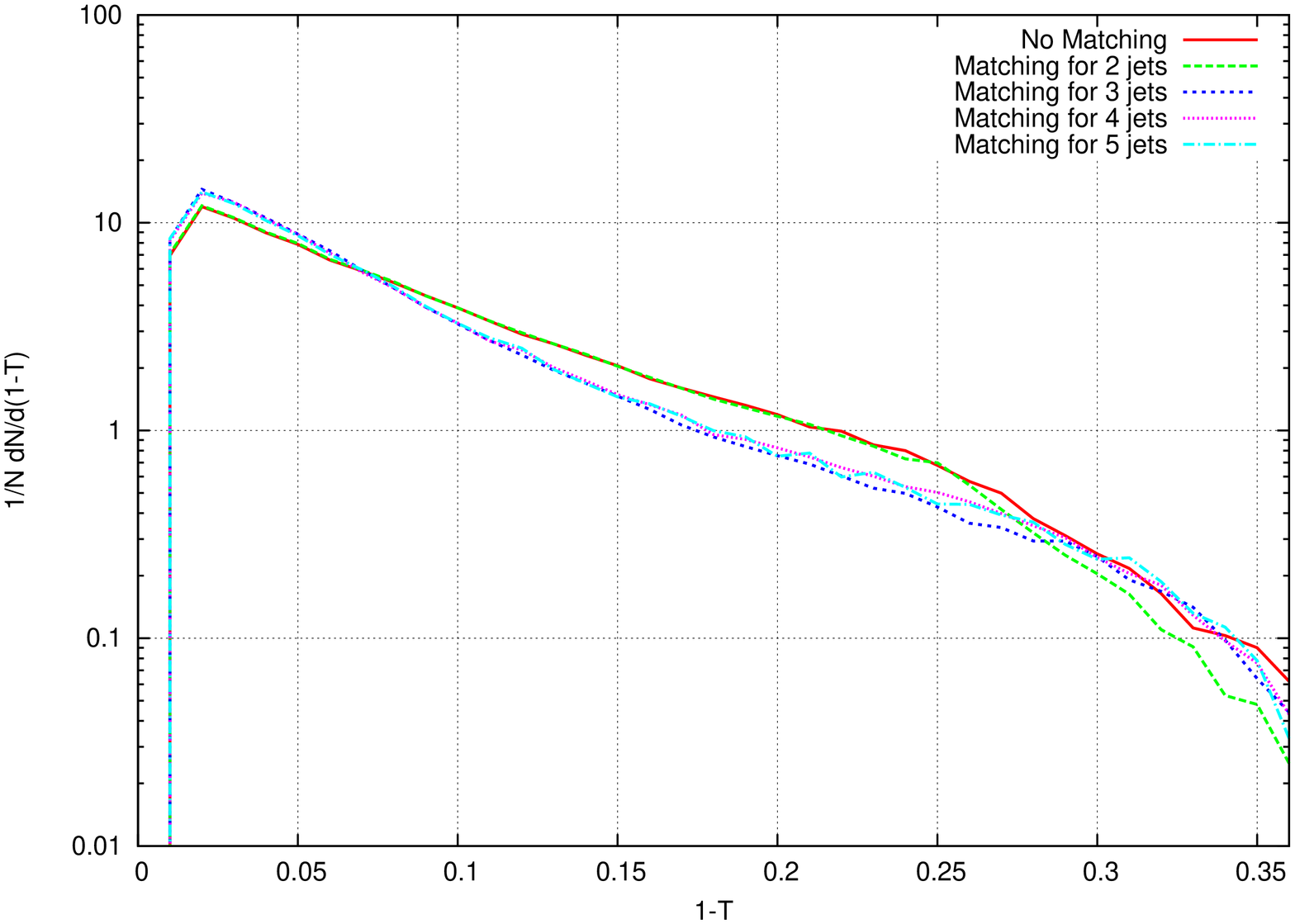}
 \includegraphics[bb=50 50 554 770,scale=0.45, angle=-90]{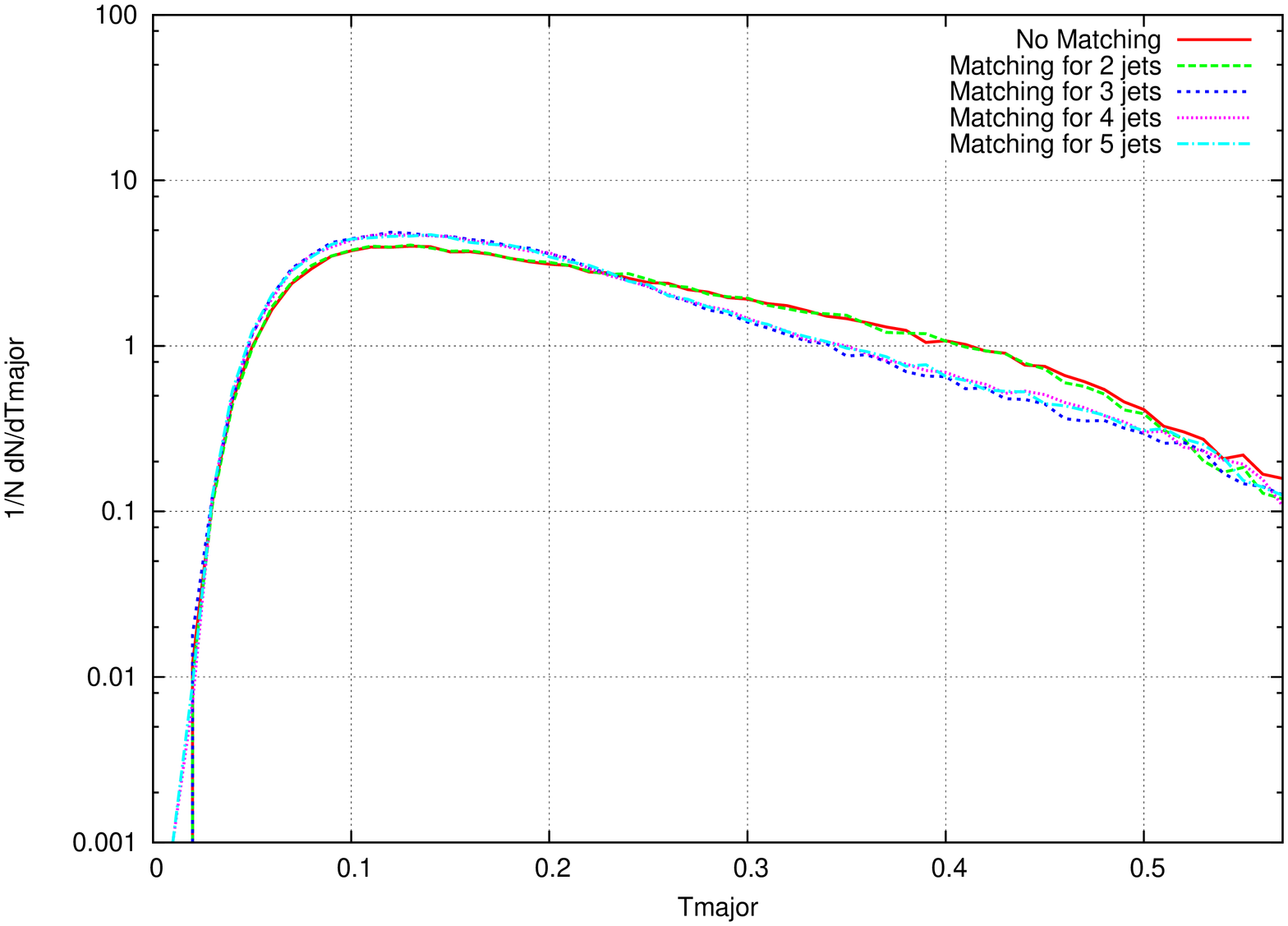}
 \caption{Plots for thrust $T$ and thrust major $T_{major}$ (\whizard\ ME + \whizard\ PS with matching).}
 \label{fig:fsrmatchingplots1W}
\end{figure}

\begin{figure}
 \centering
 \psfrag{Tminor}[l][][1][0] {$T_{minor}$}
 \psfrag{Oblateness}[l][][1][0] {$O$}
 \psfrag{1/N dN/dTminor}[l][][1][0] {$1/N \;\dif N / \dif T_{minor}$}
 \psfrag{1/N dN/dO}[l][][1][0] {$1/N \;\dif N / \dif O$}
 \includegraphics[bb=50 50 554 770,scale=0.45, angle=-90]{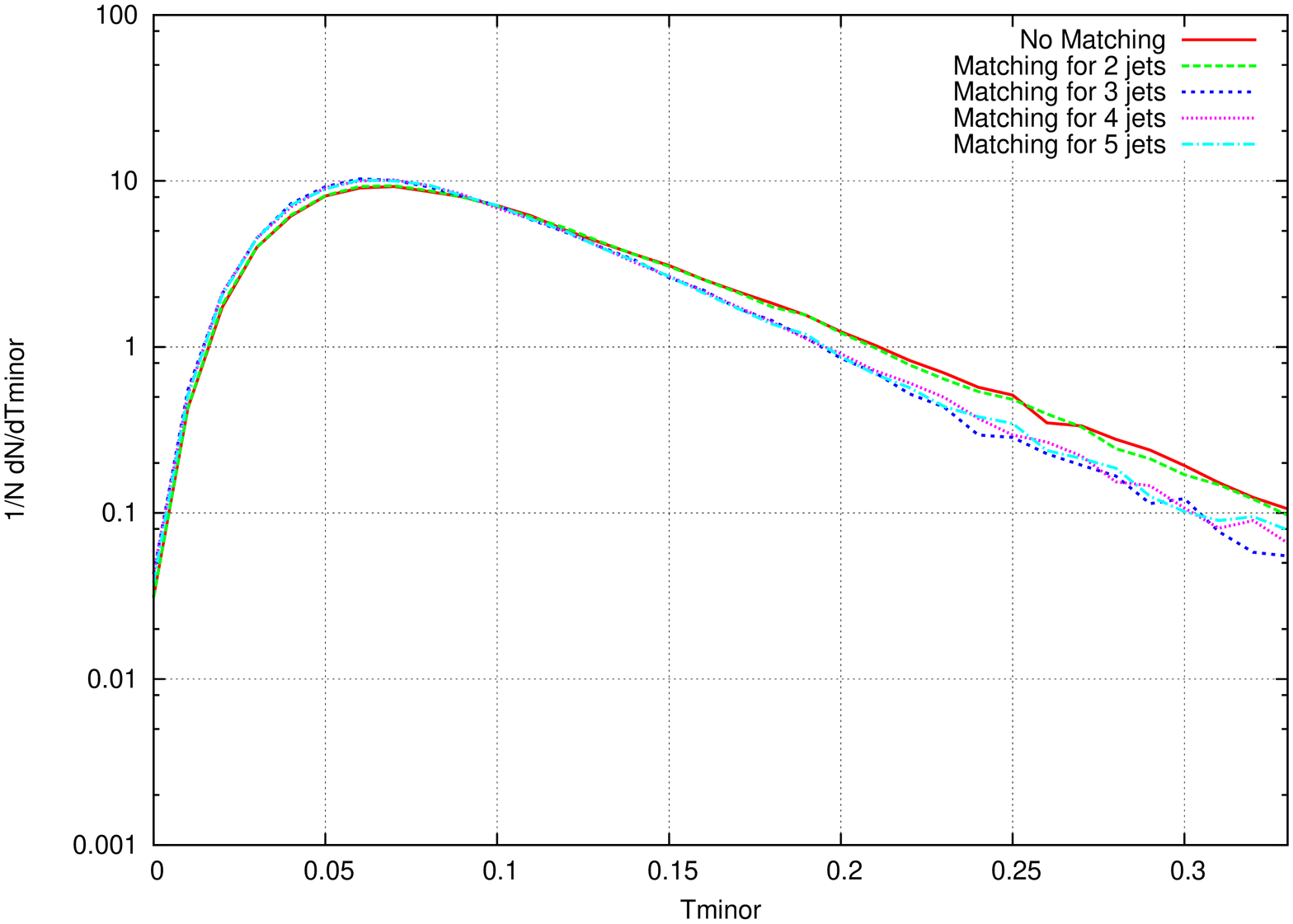}
 \includegraphics[bb=50 50 554 770,scale=0.45, angle=-90]{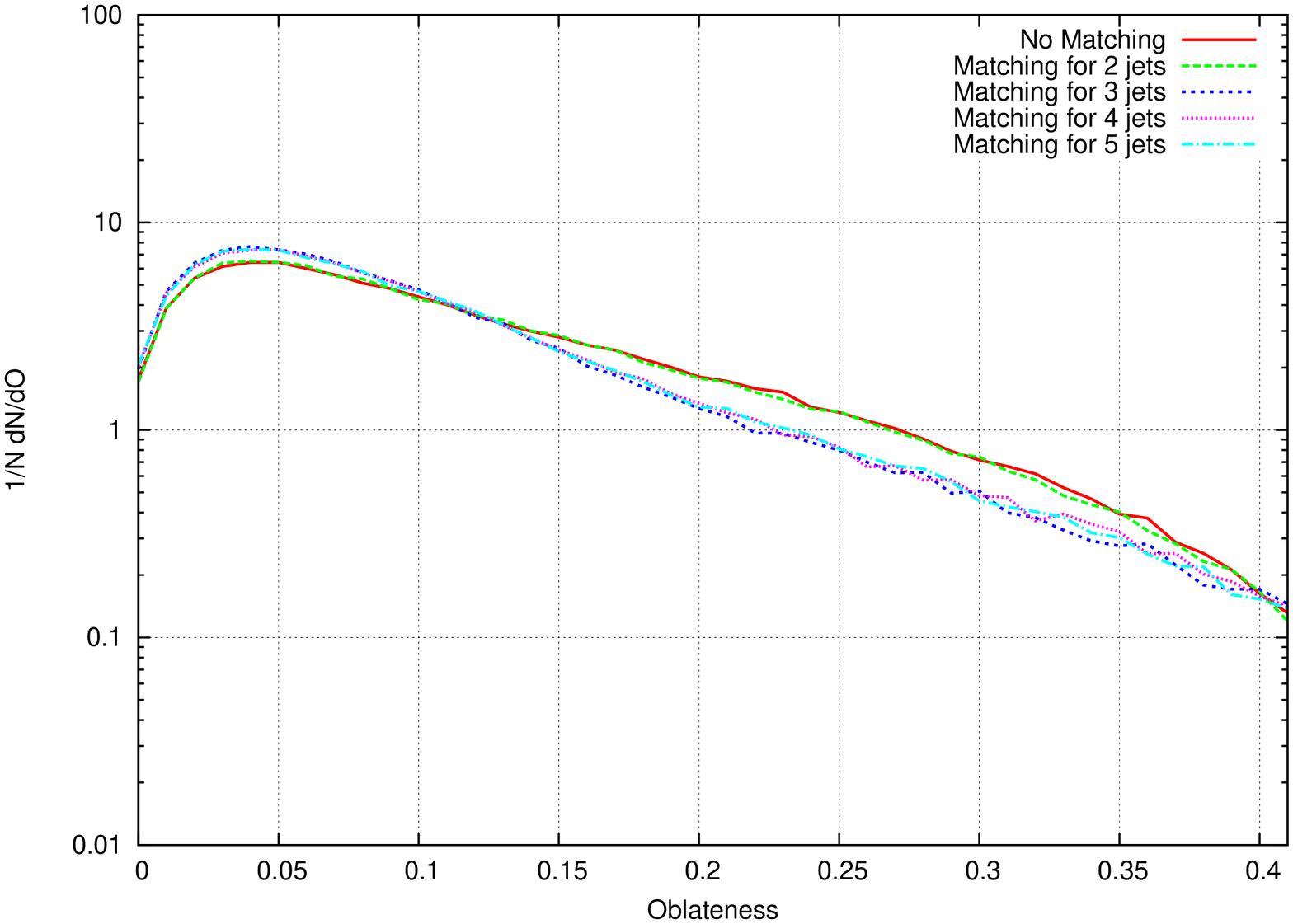}
 \caption{Plots for thrust minor $T_{minor}$ and Oblateness $O$ (\whizard\ ME + \whizard\ PS with matching).}
 \label{fig:fsrmatchingplots2W}
\end{figure}

\begin{figure}
 \centering
 \psfrag{1-T}[l][][1][0] {$1-T$}
 \psfrag{Tmajor}[l][][1][0] {$T_{major}$}
 \psfrag{1/N dN/d(1-T)}[l][][1][0] {$1/N \;\dif N / \dif (1-T)$}
 \psfrag{1/N dN/dTmajor}[l][][1][0] {$1/N \;\dif N / \dif T_{major}$}
 \includegraphics[bb=50 50 554 770,scale=0.45, angle=-90]{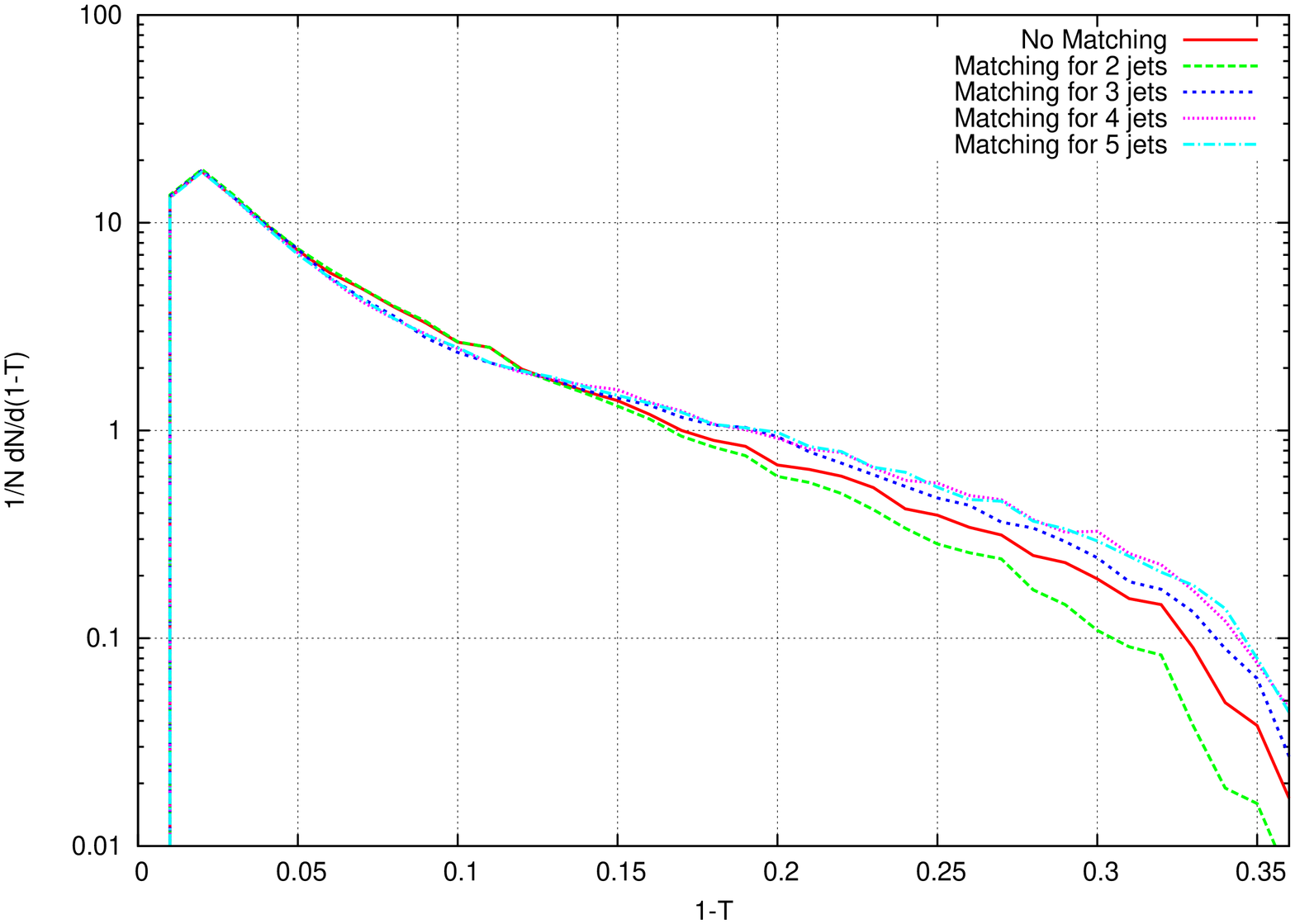}
 \includegraphics[bb=50 50 554 770,scale=0.45, angle=-90]{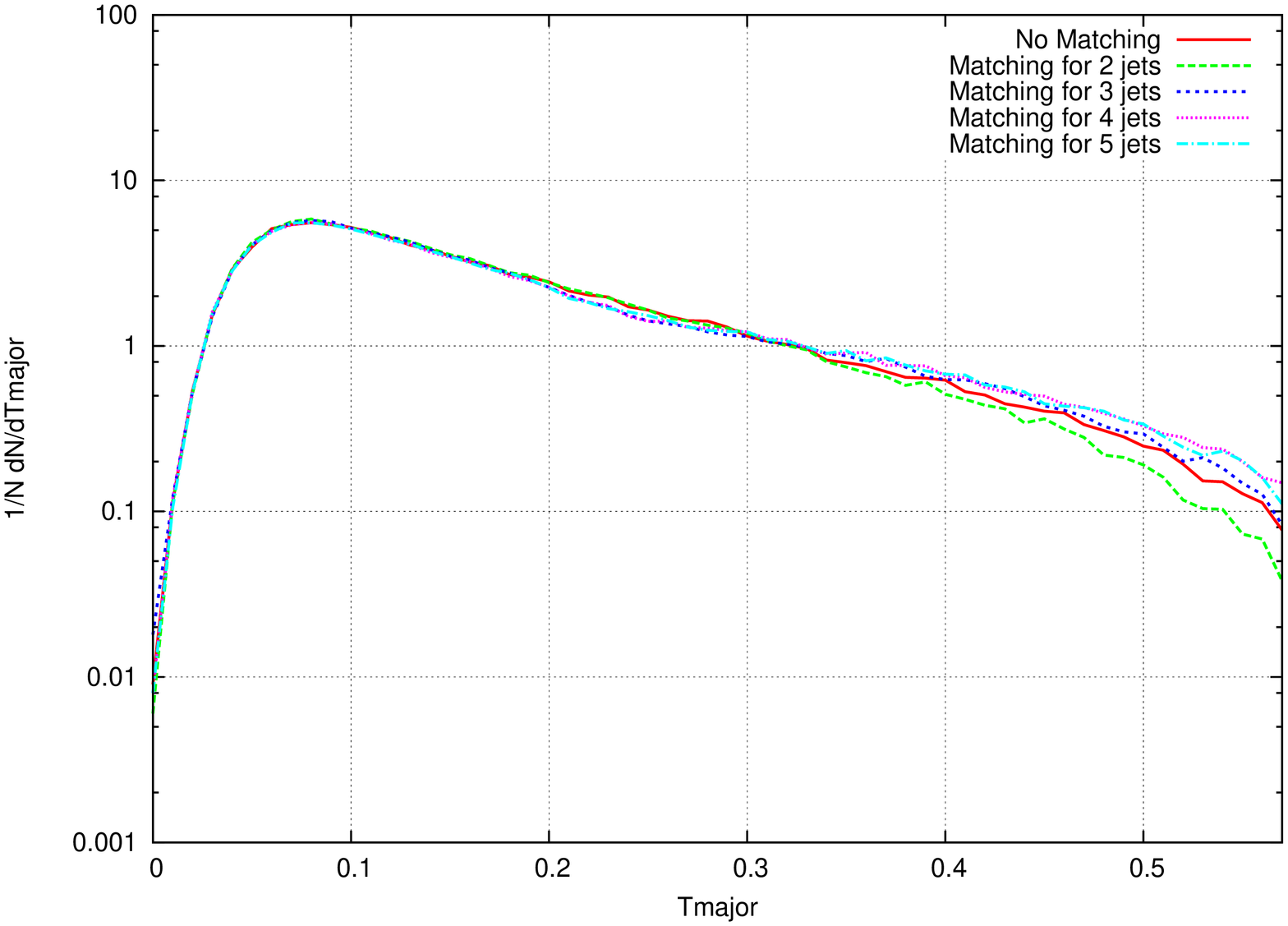}
 \caption{Plots for thrust $T$ and thrust major $T_{major}$ (\whizard\ ME + \pythia\ PS with matching).}
 \label{fig:fsrmatchingplots1P}
\end{figure}

\begin{figure}
 \centering
 \psfrag{Tminor}[l][][1][0] {$T_{minor}$}
 \psfrag{Oblateness}[l][][1][0] {$O$}
 \psfrag{1/N dN/dTminor}[l][][1][0] {$1/N \;\dif N / \dif T_{minor}$}
 \psfrag{1/N dN/dO}[l][][1][0] {$1/N \;\dif N / \dif O$}
 \includegraphics[bb=50 50 554 770,scale=0.45, angle=-90]{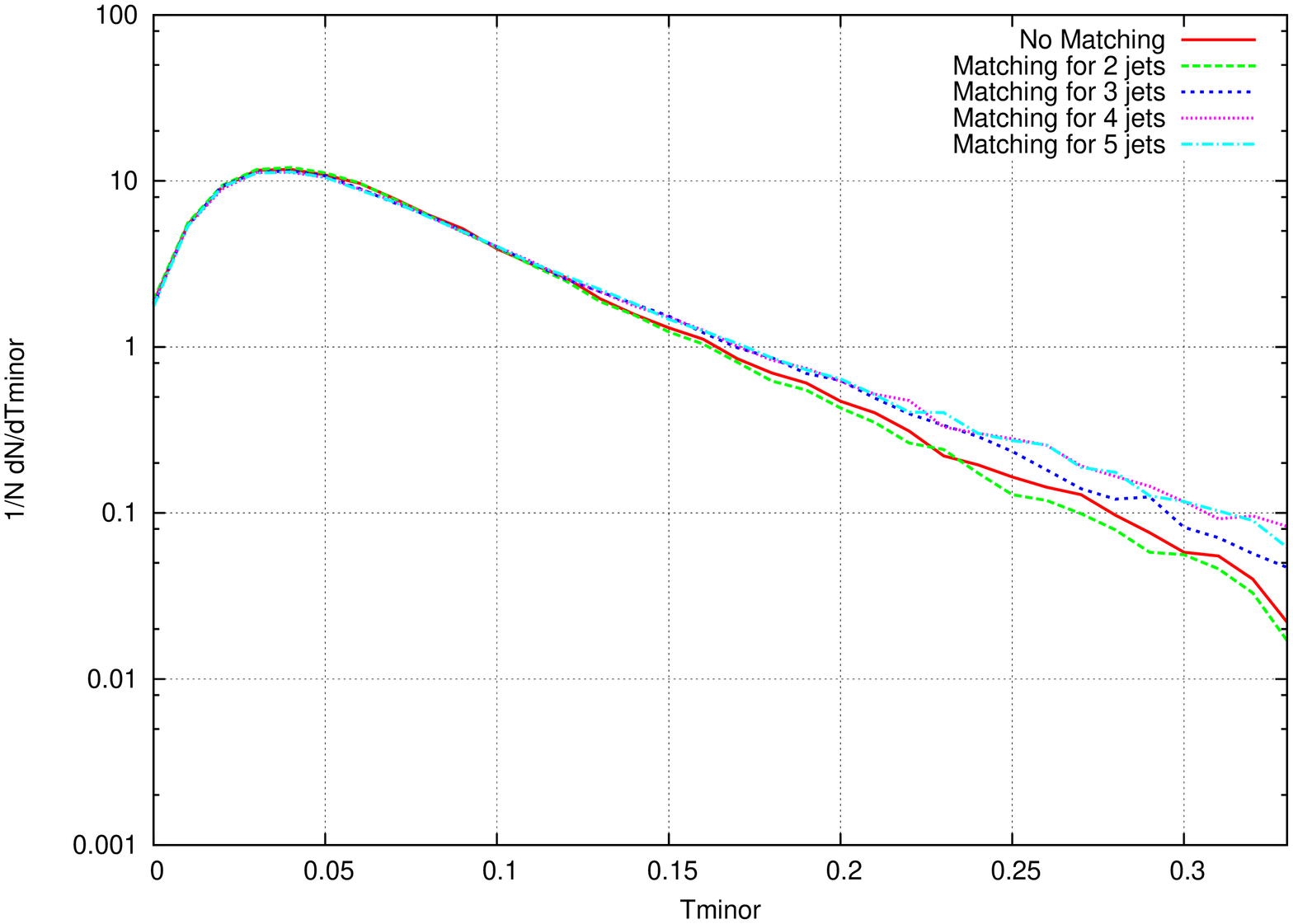}
 \includegraphics[bb=50 50 554 770,scale=0.45, angle=-90]{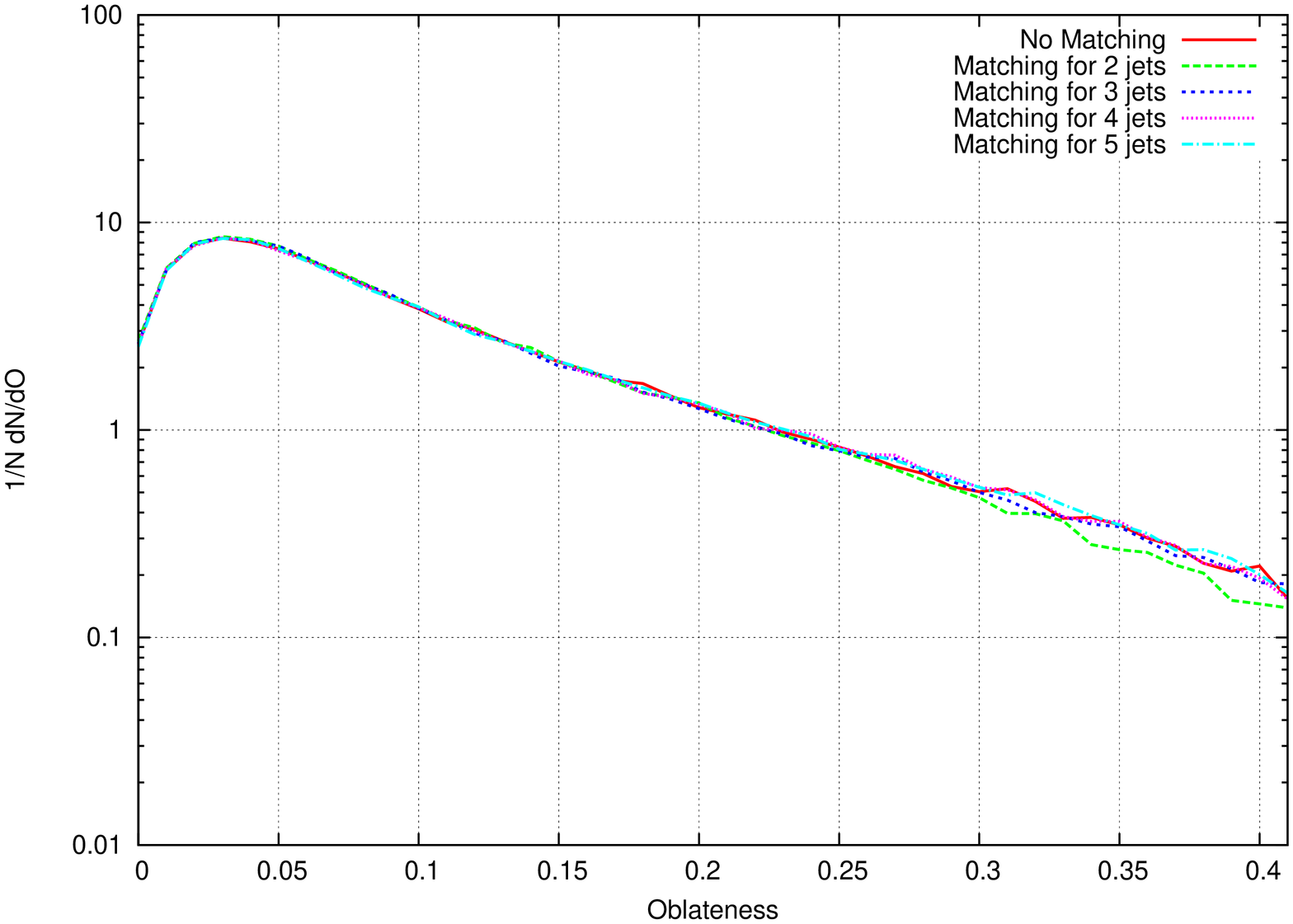}
 \caption{Plots for thrust minor $T_{minor}$ and Oblateness $O$ (\whizard\ ME + \pythia\ PS with matching).}
 \label{fig:fsrmatchingplots2P}
\end{figure}

Plots for results obtained with the MLM matching for the final-state parton shower are shown in figures \ref{fig:fsrmatchingplots1W} and \ref{fig:fsrmatchingplots2W} for events showered with \whizard's parton shower and figures \ref{fig:fsrmatchingplots1P} and \ref{fig:fsrmatchingplots2P} for \pythia. The process under consideration is $e^+ e^-\rightarrow u \bar{u}$ at a center of mass energy of $91 \,\mbox{GeV}$, hadronization was switched off. The process was simulated in five different ways, first without any matching at all and then with a variable number of additional jets from zero to three, where each additional jet could be a gluon or a $u$,$d$,$s$ or $c$ quark. For the unmatched case and for each jet multiplicity an event set consisting of 150000 events was simulated. The plots show normalized distributions for event shapes obtained from these samples.

The plots show some common features. The line for the (moot) case of no additional jets closely resembles the line for the unmatched event sample, except in the region of low thrust (right part of the upper image in figures \ref{fig:fsrmatchingplots1W} and \ref{fig:fsrmatchingplots2W}). The missing events are events where the parton shower splits a hard jet into two separated jets, so that the matching procedure can not cluster any of the two jets to the original parton and therefore rejects the event.

The lines for one, two and three additional jets lie on top of each
other so that it can be concluded that for these observables, the
inclusion of one additional jet is sufficient.  The deviations between
the unmatched and the matched event samples exhibit different
behaviour: for \pythia\ the number of spherical events is larger for
the matched sample, stemming from the better description of large
angle emissions. For \whizard\ the deviations are opposite, the number
of more pencil-like events are enhanced, while especially the number
of events with medium values of $1-T$ and $T_{maj}$ is decreased. This
can be regarded as correcting the tendency to favour more spherical
events mentioned in section \ref{sec:FSRhad}. The differences between
the distributions for matched and unmatched event samples have to be
taken into account when tuning the combination of shower and matching
to data. Therefore this can be seen as an example for an observable
which is sensitive to regions enriched by hard jet emission, and not
so much dominated by universal logarithmic terms. For such an
observable, a tuning obtained without matching cannot be reliably used
to generate matched samples. 



\begin{figure}
 \centering
 \psfrag{1-T}[l][][1][0] {$1-T$}
 \psfrag{Tmajor}[l][][1][0] {$T_{major}$}
 \psfrag{1/N dN/d(1-T)}[l][][1][0] {$1/N \;\dif N / \dif (1-T)$}
 \psfrag{1/N dN/dTmajor}[l][][1][0] {$1/N \;\dif N / \dif T_{major}$}
 \includegraphics[bb=50 50 554 770,scale=0.45, angle=-90]{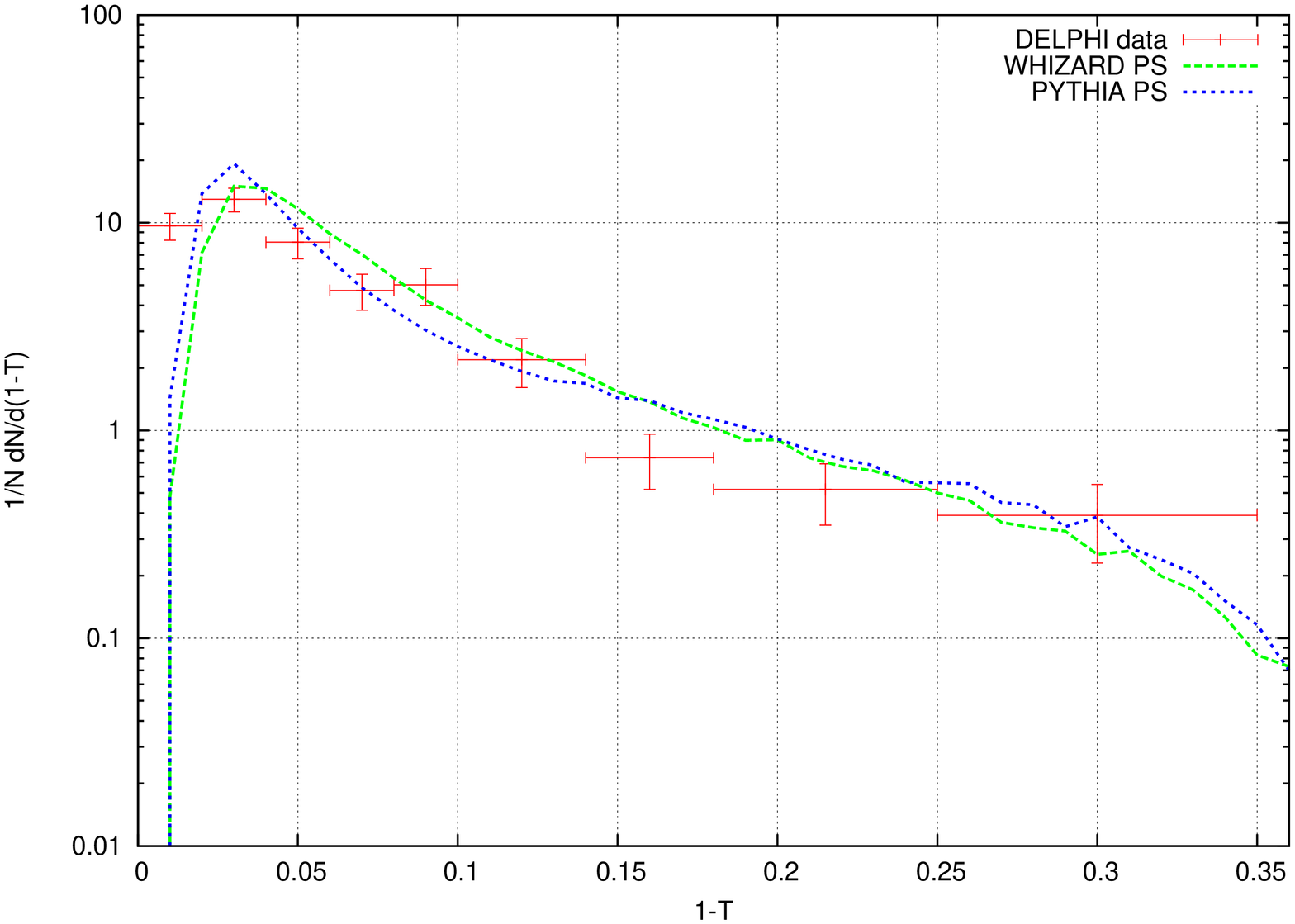}
 \includegraphics[bb=50 50 554 770,scale=0.45, angle=-90]{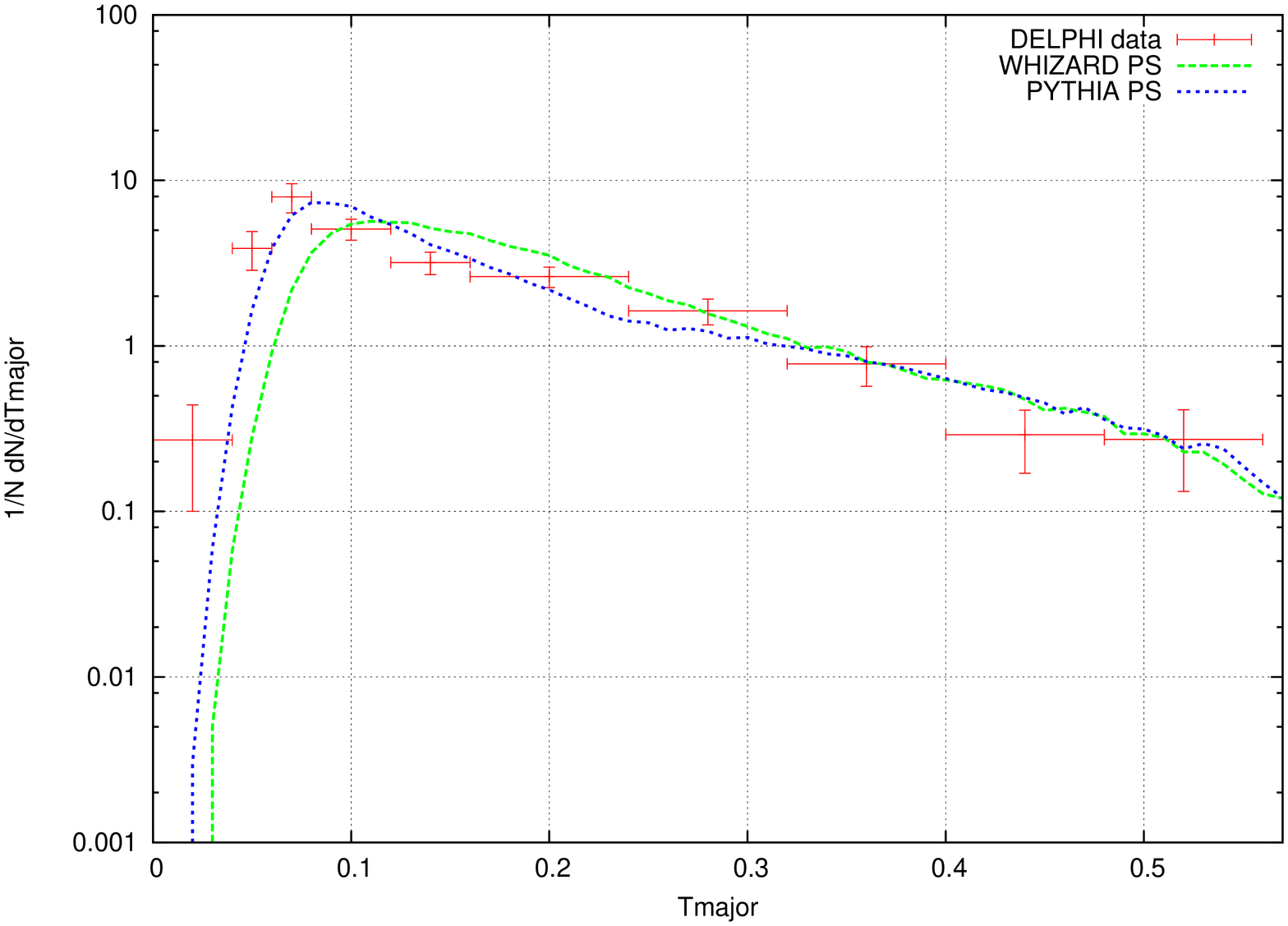}
 \caption{Comparison of the predictions for parton showers matched to the process $e^+ e^- \rightarrow u\bar{u} + 3 jets$.}
 \label{fig:fsrmatchingplotsC}
\end{figure}

We also did a comparison to data corresponding to the comparison for the unmatched showers in figure \ref{fig:fsrplotshad1}. We used the curve for the $e^+ e^- \rightarrow 5 jets$ as the sample for the matched shower. The plot is shown in figure \ref{fig:fsrmatchingplotsC}. The curve for thrust $T$ is slightly altered, most prominent differences to figure \ref{fig:fsrplotshad1} is a less pronounced peak with both showers and an increase for the \pythia\ curve for values $1-T>0.1$. The curves for Thrust major $T_{maj}$ show similar behaviour to the unmatched curves. Both reproduce the data, except for \whizard's parton shower's tendency to more spherical configurations and the small number of events in the lower $T_{maj}$-bins. Both these deficiencies have already been visible in the unmatched event samples.

\subsection{Matched Initial State Radiation}

\begin{figure}
 \centering
 \psfrag{pt}[l][][1][0] {$p_T / \mbox{GeV}$}
 \psfrag{1/N dN/dpt}[l][][1][0] {$1/N \;\dif N / \dif (p_T/\mbox{GeV})$}
 \includegraphics[bb=50 50 554 770,scale=0.45, angle=-90]{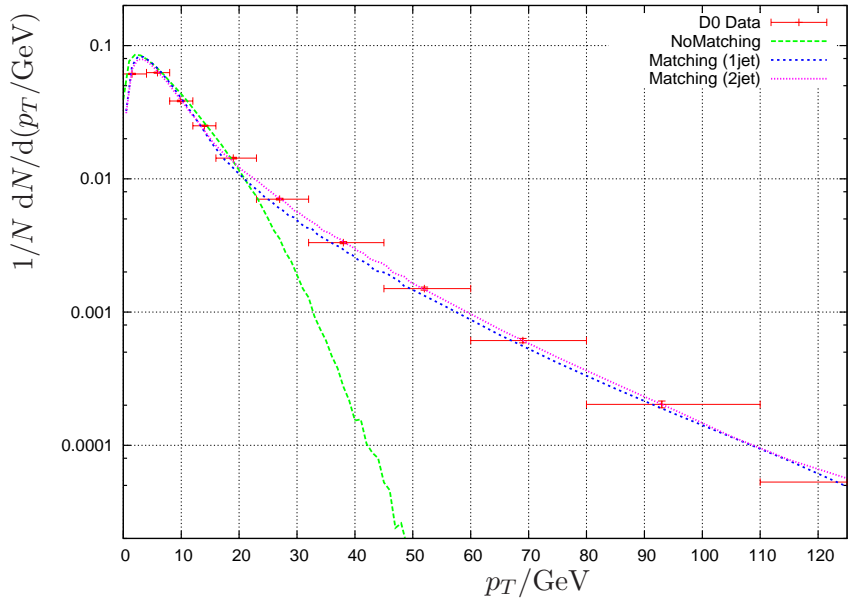}
 \caption{$Z$-Boson transverse momentum, simulated with \whizard\ ME and PS without and with matching.}
 \label{fig:isrmatchingplotW}
\end{figure}
\begin{figure}
 \centering
 \psfrag{pt}[l][][1][0] {$p_T / \mbox{GeV}$}
 \psfrag{1/N dN/dpt}[l][][1][0] {$1/N \;\dif N / \dif (p_T/\mbox{GeV})$}
 \includegraphics[bb=50 50 554 770,scale=0.45, angle=-90]{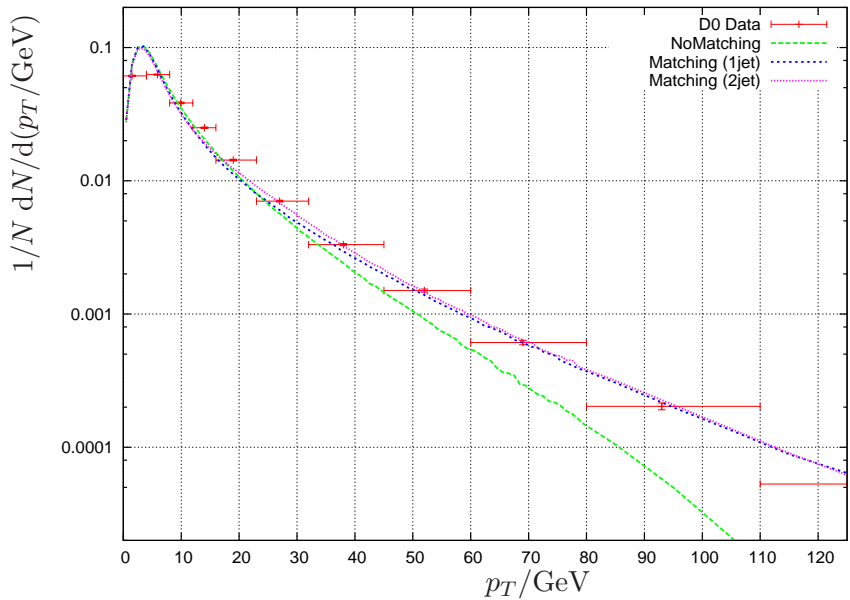}
 \caption{$Z$-Boson transverse momentum, simulated with \whizard\ ME and \pythia\ PS without and with matching for one and two additional jets.}
 \label{fig:isrmatchingplotP}
\end{figure}
\begin{figure}
 \centering
 \psfrag{pt}[l][][1][0] {$p_T / \mbox{GeV}$}
 \psfrag{1/N dN/dpt}[l][][1][0] {$1/N \;\dif N / \dif (p_T/\mbox{GeV})$}
 \includegraphics[bb=50 50 554 770,scale=0.45, angle=-90]{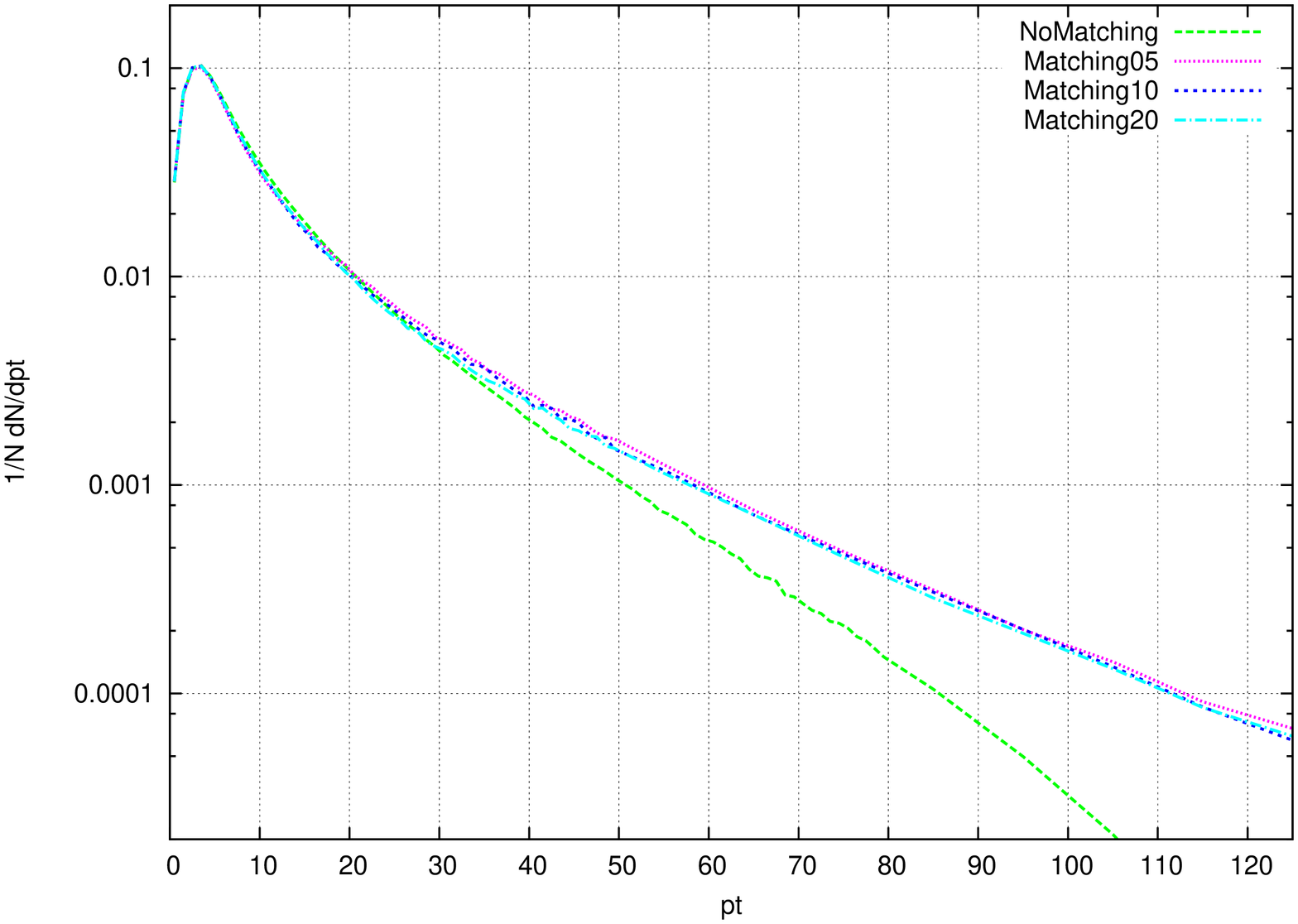}
 \caption{The ``NoMatching'' curve and the ``Matching (1jet)''(cf. figure \ref{fig:isrmatchingplotP}) curve for three different values of $p_{T\,min}$, $5\ei{GeV}$, $10\ei{GeV}$ and $20\ei{GeV}$. }
 \label{fig:isrmatchingplotP2}
\end{figure}

To test the matching procedure for the initial state we simulated the process $q \bar{q} \rightarrow Z$ and the additional corrections $j\, j \rightarrow Z\, j\,(j)$ for one (two) additional jets, $j=u,\bar{u},d,\bar{d},s,\bar{d},c,\bar{c},g$. The resulting distributions for the $Z$ boson transverse momentum are given in figure \ref{fig:isrmatchingplotW} for \whizard\ and figure \ref{fig:isrmatchingplotP} for \pythia. For comparison the measured distribution from \texttt{D0} \cite{Abazov:2010kn} was included. Note that all simulated distributions were obtained with disabled primordial $k_T$.

As expected, the results for \pythia\ do not depend much on the application of matching as its power shower approach already generates a $p_T$-distribution close to the correct distribution \cite{Miu:1998ju}\footnote{\pythia's own matching was disabled during this simulation.}. The plot for \whizard\ shows the expected addition of high-$p_T$ events, the concavity is weakend. Adding a second jet described by the matrix element does change the distribution only marginally for both showers.

Figure \ref{fig:isrmatchingplotP2} shows the dependence of the $p_T$-spectrum on the MLM-matching parameter $p_{T\,min}$. The distribution should be independent of $p_{T\,min}$, however a small difference is visible in the range $10\ei{GeV} \lesssim p_T \lesssim 80\ei{GeV}$. The high-$p_T$-tail remains stable when changing $p_{T\,min}$, the shape at the peak does not change as well. The differences are within the expected dependence on the matching parameters.

\begin{figure}
 \centering
 \psfrag{pt}[l][][1][0] {$p_T / \mbox{GeV}$}
 \psfrag{1/N dN/dpt}[l][][1][0] {$1/N \;\dif N / \dif (p_T/\mbox{GeV})$}
 \includegraphics[bb=50 50 554 770,scale=0.45, angle=-90]{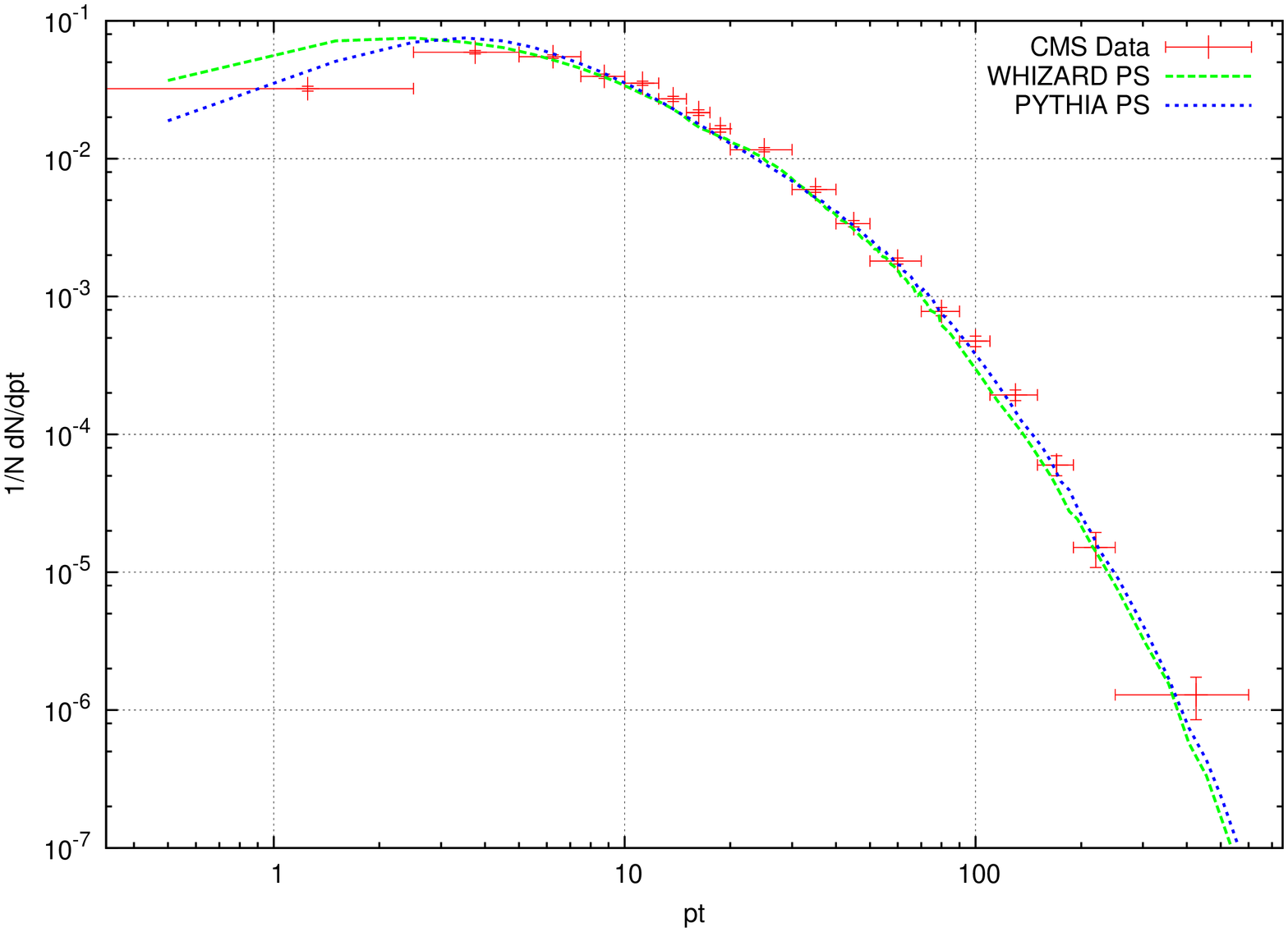}
 \caption{$Z$-Boson transverse momentum, simulated with \whizard\ ME and PS and \whizard\ ME + \pythia\ PS. A similar figure with the same data appeared in \cite{CMS}. The dashed/green/bright line is \whizard, the dotted/blue/dark line is \pythia.}
 \label{fig:isrmatchingplotLHC}
\end{figure}

As a further test, we compared the $Z$-Boson $p_T$ at the LHC. We used the recently published measurement by CMS \cite{CMS}. Except for the change from proton-antiproton beams to proton-proton beams and the increased center of mass energy $\sqrt{s}= 7 \ei{TeV}$, all other settings were the same as for the Tevatron simulation. This holds particularly for the chosen \pythia\ tune, that was obtained from measurements at Tevatron and usage at the LHC cannot be regarded as trustworthy. Nevertheless, the data can be reproduced very well, except for an overshoot in the lowest bins. As for \whizard\ there are no available tunes yet, so the dependency on a particular tune is not an issue. Note that the simulation was done with primordial $k_T$ disabled, so that the lowest bins are expected to be overpopulated. Apart from this difference, the simulation using \whizard's hard interaction, parton shower and matching procedure reproduces the data as good as the simulation performed using \whizard's hard interaction and matching, but \pythia's parton shower.

\section{Conclusion and Outlook}
\label{sec:conclusion}

In this paper, we presented an algorithm for an analytic parton
shower for both initial and final-state radiation. While this parton
shower algorithm for the final state has been known for quite some
time, the extension to the initial state had not been done up to now. 

Analytic parton showers are especially interesting for conceptual
development in a theoretical description of QCD in a hadron (but also
hadronic final state lepton) collider environment, as they allow to
determine the corresponding shower weights from the complete shower
histories. As there are also veto mechanism for probabilistic showers,
such a task is not viable there. The knowledge of complete shower
histories and weights enables one to e.g. change the hard scattering
matrix element or the PDFs and reweight the showered events to the new
hard scattering process. Furthermore, analytic parton showers might
offer the possibility to determine systematic uncertainties from a
parton shower approximation in a reliable and theoretically
well-defined way\footnote{A complementary approach has recently been presented in \cite{Giele:2011cb}. There the probability for an event is not calculated analytically. Instead a veto algorithm is used that keeps track of all accepted and vetoed branchings. For an event generated with unit weight in one setup, its weight in a different setup is given by the ratio of the joint probabilities to accept all accepted branchings while vetoing all vetoed branchings.}

Also, it might be achievable -- using analytic parton
showers -- to systematically construct higher-order corrections to
the parton shower approximation. 

The important point for a successful algorithm for an analytic
initial-state parton shower is the scale choice, specifically the
starting condition of the backward shower evolution, together with the
prescriptions for energy and momentum projections in the splittings. 

We also improved on the original algorithm for the final-state shower,
where e.g. running couplings constants within the shower evolution had
not been taken into account. For the description of complete
kinematical distributions at hadron colliders, including the
high-energy tails, we refrained from the power-shower concept, where
also hard and/or non-collinear jets are being produced by means of the
shower. Instead we use an MLM-type matching of the analytic parton shower with
matrix elements containing one or more additional hard jets
explicitly.

Together with the development of the algorithm, we made a thorough
comparison, where for now we restricted ourselves to the
parton shower from \pythia. We also made  
an extended validation of our parton shower algorithm with jet and
event shape data from the LEP experiments, from the Tevatron Run II
measurements as well as first results from the 2010/11 LHC run. 
For this task we integrated an implementation of our parton shower
algorithm into the event generator \whizard, while
hadronization needed for the comparison, is performed by means of
external packages. Our setup allows for a direct comparison of the
\pythia\ and our own parton shower using the same hard matrix
elements within the framework of the \whizard\ generator.

Without performing an overly sophisticated tuning
of the shower, we reproduced the gross features of a big number of jet
and event shape variables at lepton and hadron colliders and found in
all cases good agreement.

This paper serves as a proof of concept that an analytic parton
shower for the initial state is viable to describe QCD in a realistic
collider environment. Future lines of developments will contain a more
extensive tuning and validation of the shower as well as the matching
and merging prescription. We will also be investigating a possible
exchange of the evolution variable for the transverse momentum, $p_T$,
which would guarantee angular ordering and color coherence right from
the beginning, which might simplify or even improve on the parton
shower description given in our algorithm. A development of an
interleaved multiple interaction algorithm connected with a properly
color-connected analytic initial-state parton shower together with
its implementation is in preparation and will be part of a future
publication.

\section*{Acknowledgements}

This project has been partially supported by the Ministery of Culture 
and Science (MWK) of the state Baden-W\"urttemberg and by the Helmholtz Alliance ``Physics at 
the Terascale''. S.S. and D.W. acknowledge 
support from the Graduate School GRK1102 ``Physics at Hadron Colliders''
of the German Research Council (DFG) as well as the Scottish Physics 
Universities Alliance, SUPA.

W.K. would like to thank T. Stelzer, and S. Willenbrock for the 
hospitality at the University of Champaign-Urbana where part of this 
work was initiated. S.S. would like to thank S. Pl\"atzer for his comments on a draft of the paper.

\bibliographystyle{JHEP}
\bibliography{paper}

\clearpage
\appendix
\section{Definitions of Observables}
\label{sec:definitions}

\subsection{Event shapes}
The summations are always over all final state partons.
\begin{itemize}
\item Thrust $T$:
        \begin{displaymath}
                T = \underset{\vec{n}}{\mbox{max}} \frac{\sum_i | \vec{p}_i \cdot \vec{n}|}{\sum_i |\vec{p}_i|},
        \end{displaymath}
\item Thrust major $T_{major}$:
        \begin{displaymath}
                T_{major} = \underset{\vec{n}, \vec{n}\cdot\vec{n}_{T}=0}{\mbox{max}} \frac{\sum_i | \vec{p}_i \cdot \vec{n}|}{\sum_i |\vec{p}_i|},
        \end{displaymath}
	with the thrust axis $\vec{n}_T$.
\item Thrust minor $T_{minor}$:
        \begin{displaymath}
                T_{minor} = \qquad \frac{\sum_i | \vec{p}_i \cdot \vec{n}|}{\sum_i |\vec{p}_i|},
        \end{displaymath}
	with $\vec{n}$ perpendicular to the thrust axis $\vec{n}_T$ and the thrust major axis $\vec{n}_{T_{major}}$.
\item Oblateness $O$:
	\begin{displaymath}
		O = T_{major} - T_{minor}
	\end{displaymath}
\item Hemisphere broadenings $B$:
        \begin{eqnarray*}
         && B_\pm = \dfrac{ \sum\limits_{\pm \vec{p}_i \cdot \vec{n}_{Thrust}>0} \left| \vec{p}_i \times \vec{n}_{Thrust} \right| }{ 2 \sum\limits_i \left|\vec{p}_i \right| } \\
        && B_{max} = \mbox{max} (B_+, B_-) \qquad B_{min} = \mbox{min} (B_+, B_-) \\
        && B_{sum} = B_+ + B_- \qquad\qquad B_{diff} = \left| B_+ - B_-\right|
        \end{eqnarray*}
\end{itemize}

\subsection{Jet rates}
\label{sec:jetrates}

Jet algorithms are tools to organize the plethora of particles produced in a collision. This is done by grouping ``similar'' particles into one pseudo-particle called \emph{jet}. The criteria can be the closeness in the geometry of the detector, leading to \emph{cone-jet algorithms}, where all particles within a cone of ``radius'' $R$ are assumed to be one jet. The measure $R$ is given by $R=\sqrt{ (\Delta \eta)^2 + (\Delta \phi)^2 }$ with the pseudo-rapidity $\eta$ and the azimuthal angle $\phi$. For a further discussion of the problems arising from this approach see e.g. SISCone \cite{Salam:2007ce}. A different approach is to sequentially remove one particle after another. The procedure is to find the minimum value of the jet separations $y_{ij}, y_{ib}$ where $i$ and $j$ denote the particles and then, if the smallest value is a $y_{ij}$ both particles are removed and replaced by a combination of the two particles. If the smallest value is $y_{ib}$, the particle $i$ is removed and implicitly clustered to the beam axis. The values $y_{ij}, y_{ib}$ are given by
\begin{eqnarray}
	y_{ij} &=& 2 \mbox{min}\left( E_i , E_j \right)^2 \left( 1 - \cos\theta_{ij} \right).\nonumber \\
	y_{ib}  &=& 2 E_i^2 \left( 1 - \cos\theta_{i\,beam} \right).
	\label{eq:ktmeasure}
\end{eqnarray}
For hadronic collisions another popular definition is 
\begin{eqnarray}
	y_{ij} &=& \left( \Delta R_{ij} \right)^2 \mbox{min}\left( p_{\perp\,i}^2, p_{\perp\,j}^2 \right) \nonumber \\
	y_{ib} &=& p_{\perp\,i}^2.
\end{eqnarray}
By consecutively applying this prescription, every event can be gradually clustered to a $2\rightarrow 2$ process. For each step of the clustering, the $y$ value of the last clustering gives the jet separation for the corresponding number of jets. These prescriptions for the distance-measures compose the so-called \emph{$k_T$-algorithm} \cite{Catani1993187}. The algorithm can be varied by replacing the 2 in the exponent of $\left( \Delta R_{ij} \right)$ in equation \eqref{eq:ktmeasure}. Other values that have been studied are 0 and $-2$, changing to 0 leads to the \emph{Cambridge-Aachen} algorithm, while changing to $-2$ produces the \emph{anti-$k_T$-algorithm}.

\end{document}